# Structural, elastic, electronic, bonding, thermal, and optical properties of topological Weyl semimetal Ta$X$ ($X$ = P, As): Insights from *ab-initio* calculations

M. I. Naher, S. H. Naqib*

*Department of Physics, University of Rajshahi, Rajshahi 6205, Bangladesh*
*Corresponding author email: salehnaqib@yahoo.com

**Abstract**
In recent days, study of topological Weyl semimetals have become an active branch of physics and materials science because they led to realization of the Weyl fermions and exhibited protected Fermi arc surface states. Therefore, topological Weyl semimetals Ta$X$ ($X$ = P, As) are important electronic systems to investigate both from the point of view of fundamental physics and potential applications. In this work, we have studied the structural, elastic, mechanical, electronic, bonding, acoustic, thermal and optical properties of Ta$X$ ($X$ = P, As) in detail via first-principles method using the density functional theory. A comprehensive study of elastic constants and moduli shows that both TaP and TaAs possesses low to medium level of elastic anisotropy (depending on the measure), reasonably good machinability, mixed bonding characteristics with ionic and covalent contributions, brittle nature and relatively high Vickers hardness with a low Debye temperature and melting temperature. The minimum thermal conductivities and anisotropies of Ta$X$ ($X$ = P, As) are calculated. Bond population analysis supports the bonding nature as predicted by the elastic parameters. The bulk electronic band structure calculations reveal clear semi-metallic features with signature Dirac cone-like dispersions near the Fermi level. A pseudogap in the electronic energy density of states at the Fermi level separating the bonding and the antibonding states indicates significant electronic stability of tetragonal Ta$X$ ($X$ = P, As). The reflectivity spectra show almost non-selective behavior over a wide range of photon energy encompassing visible to mid-ultraviolet regions. High reflectivity over wide spectral range makes Ta$X$ suitable as reflecting coating. Ta$X$ ($X$ = P, As) are very efficient absorber of ultraviolet radiation. Both the compounds are moderately optically anisotropic owing to the anisotropic nature of the electronic band structure. The refractive indices are very high in the infrared to visible range. All the energy dependent optical parameters show metallic features and are in complete accord with the underlying bulk electronic density of states calculations.

# 1. Introduction

Topological quantum materials, such as topological insulators [1, 2], semimetals (Dirac or Weyl) [3-6], metals [7, 8], and superconductors [9, 10] have recently become a subject of great interest in basic science and technology, both theoretically and experimentally. Although the topological system starts with the remarkable discovery of topological insulators, recently the significant focus has been shifted towards topological Dirac or Weyl semimetals (WSM). Unlike ordinary metals (where most of their electronic properties are related to the existence of a Fermi surface), the main focus of WSMs is on the appearance of band touching points or nodes at the Fermi energy, where two or more bands are exactly degenerate at particular values of the crystal momentum within the Brillouin zone (BZ). Basically, WSMs are related to Dirac semimetals (DSMs), which can be realized by breaking either the time reversal symmetry or the spatial inversion symmetry of Dirac semimetals. In this situation the Dirac point splits into two Weyl points of opposite chirality [11–13]. The WSM is an unusual solid that hosts the Weyl fermions. Weyl fermions are massless chiral fermions with a relativistic spin of 1/2 that are allowed solutions of Dirac's equation [14–16]. Weyl fermions are relevant to both high energy particle physics and low-energy in condensed matter physics and play a crucial role in quantum field theory. The specific features of WSMs are the unique bulk Weyl fermions, the surface Fermi-arcs (open Fermi surfaces), chiral anomaly, drum-head surface states, gapless in the bulk and a pair of Weyl points (band crossing points) existing with opposite parity. All these exotic electronic features lead to many intriguing properties such as extremely large (sometimes negative) magneto-resistance, chiral magnetic effects, high carrier mobility, light effective mass, nontrivial Berry phase, quantum anomalous Hall effect, and novel quantum oscillations [17-21]. These properties make WSMs interesting for the future applications in electronic, spintronic and quantum computing devices.

TaAs and TaP are newly predicted [22, 23] and experimentally synthesized [24-28] topological Weyl semimetals. TaP and TaAs are Type-I topological Weyl semimetals with naturally broken inversion symmetry [23]. Several physical properties of TaP and TaAs have been studied both theoretically and experimentally so far [23, 29-40]. Among those elastic, thermo-physical and optical properties of isostructural Ta$X$ ($X$ = P, As) have not been explored in details. For example, analysis of the Cauchy pressure, tetragonal shear modulus, linear compressibility, internal strain parameter, machinability index, hardness, and anisotropy in elastic moduli are still lacking. The thermo-physical properties such as Debye temperature, melting temperature and minimum thermal conductivity have not been investigated thoroughly yet. The bonding nature is also unexplored. Besides energy dependence of the optical constants are still unknown. A thorough understanding of the elastic, mechanical, thermal, bonding and optical response of these compounds is necessary to unravel the potential of the titled WSMs for possible applications. This constitutes the primary motivation of the present study.

The rest of this manuscript has been structured as follows: In Section 2, we briefly described the computational methodology. In Section 3, we have presented the results of our computations and their analyses. Finally, the important features of this study are summarized and concluded in Section 4.

## 2. Computational methodology

All the structural, elastic, electronic and optical properties calculations were carried out by employing plane wave pseudopotential approach based on the density functional theory (DFT) [41, 42] implemented in the CASTEP (CAmbridge Serial Total Energy Package) simulation code [43]. The electronic exchange-correlation energy has been selected using the generalized gradient approximation (GGA) of Perdew-Burke-Ernzerhof (PBE) scheme [44].The coulomb potential energy caused by the interaction between the valence electrons and ion cores has been modeled by the Vanderbilt-type ultra-soft pseudopotential [45]. Use of ultra-soft pseudopotential saves us substantial computational time with little loss of computational accuracy. To perform pseudo atomic calculations, the following valence electron configurations have been considered: $5d^3 6s^2$ for Ta, $3s^2 3p^3$ for P and $4s^2 4p^3$ for As atoms, respectively. To get the lowest energy crystal structure of Ta$X$, geometry optimization was performed using the Broyden–Fletcher–Goldfarb–Shanno (BFGS) minimization scheme [46]. The cut off energy for the plane-wave expansion for both the compounds were set at 400 eV. The sampling of Brillouin zone (BZ)was carried out using the Monkhorst Pack mesh [47] with a mesh size of 11x11x9 and 10x10x8 $k$-points for TaAs and TaP, respectively. Geometry optimization of both TaP and TaAs were performed using total energy convergence tolerance of $10^{-5}$ eV/atom, maximum lattice point displacement within $10^{-3}$Å, maximum ionic Hellmann-Feynman force within 0.03 eVÅ$^{-1}$ and maximum stress tolerance of 0.05 GPa, with finite basis set corrections [48]. These selected levels of tolerances produced reliable estimates of structural, elastic and electronic band structure properties with an optimum computational time.

The single crystal elastic constants, $C_{ij}$, for tetragonal structure were calculated based on stress-strain method [49]. From symmetry considerations, a tetragonal crystal has six independent elastic constants ($C_{11}$, $C_{33}$, $C_{44}$, $C_{66}$, $C_{12}$ and $C_{13}$).All the others elastic properties, such as the bulk modulus ($B$) and shear modulus ($G$), can be evaluated from the calculated values of single crystal elastic constants $C_{ij}$ by using the Voigte-Reusse-Hill (VRH) approach [50, 51].

The imaginary part of the dielectric function, $\varepsilon_2(\omega)$, has been calculated by using the CASTEP supported formula,

$$\varepsilon_2(\omega) = \frac{2e^2\pi}{\Omega\varepsilon_0} \sum_{k,v,c} |\langle \Psi_k^c|\hat{u}.\vec{r}|\Psi_k^v\rangle|^2 \ \delta(E_k^c - E_k^v - E) \qquad (1)$$

where, $\Omega$ is the unit cell volume, $\omega$ is the frequency of the incident photon, $e$ is the charge of an electron, $\hat{u}$ is the unit vector defining the incident electric field polarization, and $\Psi_k^c$ and $\Psi_k^v$ are the conduction and valence band wave functions at a given wave-vector *k*, respectively. This formula makes use of the calculated electronic band structure. The real part of the dielectric function, $\varepsilon_1(\omega)$, has been found from the corresponding imaginary part $\varepsilon_2(\omega)$ using the Kramers-Kronig transformation equation. Once the values of $\varepsilon_1(\omega)$ and $\varepsilon_2(\omega)$ are known, the refractive index, the absorption coefficient, the energy loss-function, the reflectivity, and the optical conductivity can be extracted from them [52].

The Mulliken bond population analysis [53] has been used widely to understand the bonding characteristics of solids. For TaP and TaAs we have used a projection of the plane-wave states onto a linear combination of atomic orbital (LCAO) basis sets [54, 55]. The Mulliken bond population analysis can be implemented using the Mulliken density operator written on the atomic (or quasi-atomic) basis:

$$P_{\mu\nu}^M = \sum_{g'}\sum_{\nu'} P_{\mu\nu'}(g')S_{\nu'\nu}(g-g') = L^{-1}\sum_{k} e^{-ikg}(P_k S_k)_{\mu\nu'} \quad (2)$$

and a charge on atom *A* is defined as,

$$Q_A = Z_A - \sum_{\mu \in A} P_{\mu\mu}^M \quad (3)$$

where $Z_A$ is a charge of nucleus or atomic core (in simulations using the atomic pseudo-potential).

## 3. Results and analysis
### 3.1 Structural Properties

Ta*X* (*X* = P and As) assumes body-centered tetragonal crystal structure with space group *I*4$_1$*md*, (No. 109) without inversion symmetry. Figure 1 shows the crystal structure of Ta*X*. The Ta and *X* atoms occupy the following Wyckoff position in the unit cell [23]: Ta atoms at (0,0,0) and *X* atoms at (0,0,0.416). The unit cell of Ta*X* contains four Ta atoms and four *X* atoms. Table 1 listed the results of first-principles calculations of structural properties of these materials together with their available theoretical and experimental values [33, 56-57]. The calculated values of lattice constants are in good agreement with previous results.

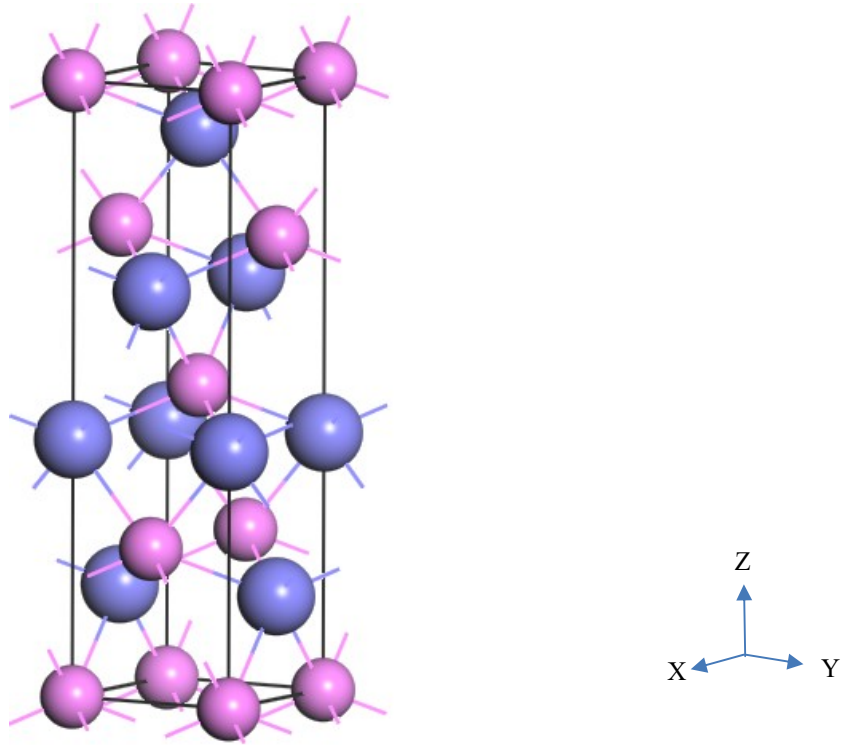

**Figure 1.** 3D Crystal structure of Ta$X$ ($X$ = P, As) unit cell. The violet atoms are Ta and the blue ones are $X$ ($X$ = P, As).

**Table 1**
Calculated and previously estimated lattice constants $a$(Å) and $c$(Å), $c/a$ ratio, equilibrium volume $V_o$(Å$^3$) and bulk modulus $B$ (GPa) of TaP and TaAs.

| Compounds | $a$ | $c$ | $c/a$ | $V_o$ | $B$ | Ref. |
|---|---|---|---|---|---|---|
| TaP | 3.37 | 11.44 | 3.39 | 129.72 | 196.49 | This |
|  | 3.32 | 11.34 | - | - | - | [56]$^{Exp.}$ |
|  | 3.32 | 11.36 | - | - | - | [57]$^{Exp.}$ |
|  | 3.34 | 11.43 | - | - | - | [33]$^{Theo.}$ |
| TaAs | 3.49 | 11.71 | 3.36 | 142.75 | 167.96 | This |
|  | 3.43 | 11.65 | - | - | - | [56]$^{Exp.}$ |
|  | 3.44 | 11.66 | - | - | - | [57]$^{Exp.}$ |
|  | 3.47 | 11.76 | - | - | - | [33]$^{Theo.}$ |

## *3.2 Mechanical and Elastic Properties*

Elastic constants of compounds are important parameters to determine mechanical stability and stiffness against an externally applied stress. Due to symmetry consideration, a crystal with tetragonal structure has six independent elastic constants: $C_{11}$, $C_{33}$, $C_{44}$, $C_{66}$, $C_{12}$ and $C_{13}$. The calculated elastic constants are listed in Table 2. For mechanical stability, according to Born-Huang conditions, a tetragonal system requires to satisfy the following inequality criteria [58]: $C_{11}>0$, $C_{33}>0$, $C_{44}>0$, $C_{66}>0$, $(C_{11}-C_{12})>0$, $(C_{11}+C_{33}-2C_{13})>0$, $\{2(C_{11}+C_{12})+C_{33}+4C_{13}\}>0$. All the elastic constants of TaP and TaAs are positive and satisfy these mechanical stability criteria. This indicates that both TaP and TaAs are mechanically stable.

The resistance to linear compression along [100] and [001] directions are characterized by $C_{11}$ and $C_{33}$, respectively. Here it is seen that for both the compounds $C_{11}$ is larger than $C_{33}$. This indicates that for both TaP and TaAs the bonding strength/compressibility along [100] direction are stronger/lesser than those along [001] direction. Also, TaAs is more compressible than TaP. The elastic constant $C_{44}$ parameterize the resistance to shear deformation with respect to a tangential stress applied to the (100) plane in the [010] direction of the compound. Here it is seen that for both TaP and TaAs, $C_{44}$ is lower than $C_{11}$ and $C_{33}$, which indicates that both the compounds are more easily deformed by a shear in comparison to a unidirectional stress along any of the three crystallographic directions. For both TaP and TaAs, $C_{44}$ is lower than $C_{66}$, which indicates that the shear along the (100) plane is easier relative to the shear along the (001) plane. Since $C_{11}+C_{12}>C_{33}$ for both the compounds, we can say that the bonding in the (001) plane is more rigid elastically than that along the *c*-axis as well as the elastic tensile modulus is higher on the (001) plane than that along the *c*-axis. The tetragonal shear modulus, $(C' = \frac{C_{11}-C_{12}}{2})$ of a crystal is the measure of crystal's stiffness (the resistance to shear deformation by a shear stress applied in the (110) plane in the $[1\bar{1}0]$ direction).

The Kleinman parameter ($\zeta$), also known as internal strain parameter, is an indicator which measures stability of a compound against stretching and bending. The Kleinman parameter ($\zeta$) is calculated using following equation [59]:

$$\zeta = \frac{C_{11} + 8C_{12}}{7C_{11} + 2C_{12}} \qquad (4)$$

The Kleinman parameter is a dimensionless parameter whose value generally lies in the range $0 \leq \zeta \leq 1$. The lower and upper limits of $\zeta$ ($\zeta = 0$ and $\zeta = 1$, respectively) represent insignificant contribution of bond bending to resist the external stress and insignificant contribution of bond stretching/contracting to resist the external applied stress, respectively. The calculated values of $\zeta$ of TaP and TaAs are 0.612 and 0.647, respectively. From which we predict that mechanical strength in both TaP and TaAs is mainly dominated by bond bending contribution over bond stretching or contracting. The Kleinman parameter also explains the relative position of the cation and anion sub-lattice under volume conserving strain distortions for which positions are not fixed by crystal symmetry [60].

The isotropic bulk modulus ($B$) and shear modulus ($G$) (by the Voigt-Reuss-Hill (VRH) method), Young's modulus ($Y$), Poisson's ratio ($v$) and hardness ($H_V$) of the compounds are calculated using following well known equations [61-63]:

$$B_H = \frac{B_V + B_R}{2} \tag{5}$$

$$G_H = \frac{G_V + G_R}{2} \tag{6}$$

$$Y = \frac{9BG}{(3B + G)} \tag{7}$$

$$v = \frac{(3B - 2G)}{2(3B + G)} \tag{8}$$

$$H_V = \frac{(1 - 2v)Y}{6(1 + v)} \tag{9}$$

For both TaP and TaAs, larger value of $B$ compared to $G$ (Table 3) indicates that the mechanical strength will be limited by the shear deformation. It is known that the large value of shear modulus indicates pronounced directional bonding between atoms [64]. The Young's modulus is the ratio between tensile stress to the tensile strain. $Y$ is the measure of the resistance (stiffness) of an elastic solid to a change in its length [65, 66] and provides with a measure of thermal shock resistance. The covalent nature of a material increases with Young's modulus [67]. The lattice thermal conductivity and Young's modulus of a material are related as: $K_L \sim \sqrt{Y}$ [68]. Pugh's ratio [69-71] provides with the information about the brittle/ductile nature of a material. If the value of $G/B$ is higher than 0.50, the material is brittle, otherwise it would be ductile. In our case, the $G/B$ values of TaP and TaAs are 0.58 and 0.55, respectively which indicate that both the compounds are expected show brittle behavior.

A Poisson's ratio, $v \sim 0.31$ is another indicator of brittle and ductile threshold [72]. This also implies that both TaP and TaAs are brittle in nature. For central-forces, the lower and upper limits of $v$ for a solid are 0.25 and 0.50, respectively [73, 74]. From the values of Poisson's ratio in Table 3, we can predict that interatomic forces of TaP and TaAs are central in nature. Poisson's ratio is also an indicator of presence of ionic and covalent bonding in a compound. For ionic and covalent materials the values of $v$ are typically 0.25 and 0.10, respectively [75]. The calculated Poisson's ratio of TaP and TaAs are 0.26 and 0.27, respectively. This implies that ionic contribution is present in both the compounds.

Cauchy pressure $(C" = C_{12} - C_{44})$ is also another indicator of brittle/ductile nature of a compound. A negative Cauchy pressure suggests brittleness while a positive one means damage tolerance and ductility of a material [76]. Cauchy pressure also describes the angular characteristics of atomic bonding in a solid [77]. Positive value of Cauchy pressure indicates ionic bonding in a material and the negative value of Cauchy pressure indicates covalent bonding in a material. Cauchy pressure of both TaP and TaAs are positive, which suggest that the compounds under consideration are ductile in nature. According to Pettifor's rule [77], materials with large positive Cauchy pressures have more metallic bonds and thus become more ductile, on the other hand, if Cauchy pressures of the materials are strongly negative, they have more angular bonds, and thus exhibit more brittleness. It should be noted that positive Cauchy pressure found for TaP and TaAs can be somewhat misleading since corrections due to many body interaction among atoms and electron gas are not taken into account in determining the elastic constants [78]. Positive value of Cauchy pressure for Ta$X$ probably implies that in addition to ionic and covalent bondings, these compounds have some metallic bondings as well.

Machinability of a material has become a useful topic in today's industry, because it is related to a number of variables like, the inherent properties or characteristics of the work materials, cutting tool material, tool geometry, the nature of tool engagement with the work, cutting conditions, type of cutting, cutting fluid, and machine tool rigidity and its capacity. The machinability index, $\mu_M$ of a material is calculated using following equation [79]:

$$\mu_M = \frac{B}{C_{44}} \qquad (10)$$

which can also be used as a measure of plasticity [80-83] and lubricating property of a material. From this equation we can say that, a compound with lower $C_{44}$ value gives better dry lubricity. Larger value of $B/C_{44}$ of a compound indicates excellent lubricating properties, lower feed forces, lower friction value, and higher plastic strain value. The $B/C_{44}$ values of TaP and TaAs are 1.78 and 1.93, respectively. This implies the presence of good level of machinability. Hardness value is also needed to understand elastic and plastic properties of a compound. The calculated values of hardness of TaP and TaAs are 18.07GPa and 14.07 GPa, respectively.

**Table 2**

Calculated elastic constants, $C_{ij}$ (GPa), Cauchy pressure, $C''$ (GPa), tetragonal shear modulus, $C'$ (GPa), Kleinman parameter ($\zeta$) for TaP and TaAs at P = 0 GPa and T = 0 K.

| Compounds | $C_{11}$ | $C_{33}$ | $C_{44}$ | $C_{66}$ | $C_{12}$ | $C_{13}$ | $C''$ | $C'$ | $\zeta$ | Ref. |
|---|---|---|---|---|---|---|---|---|---|---|
| TaP | 338.14 | 281.95 | 110.15 | 183.34 | 163.95 | 124.58 | 53.80 | 87.10 | 0.61 | This |
|  | 379.1 | 308.5 | 120.5 | 215.5 | 176.3 | 141.7 | - | - | - | [35]<sup>Theo.</sup> |
|  | 279.76 | 234.41 | 87.08 | 166.80 | 147.17 | 109.60 | 60.09 | 66.30 | 0.65 | This |
| TaAs | 310.5 | 256.1 | 94.5 | 194.5 | 164.2 | 128.9 | - | - | - | [35]<sup>Theo.</sup> |
|  | 310.3 | 260.7 | 93.0 | 197.8 | 178.0 | 138.0 | - | - | - | [34]<sup>Theo.</sup> |

**Table 3**

The calculated isotropic bulk modulus $B$ (in GPa), shear modulus $G$ (in GPa), Young's modulus $Y$ (in GPa), Pugh's indicator $G/B$, Machinability index $\mu_M$, Poisson's ratio $\nu$ and hardness $H_V$ (in GPa) of TaP and TaAs compounds deduced from Voigt, Reuss, and Hill (VRH) approximations.

| TaX | B | | | G | | | Y | $\frac{G_V}{G_R}$ | $\frac{B_V}{B_R}$ | G/B | $\mu_M$ | $\nu$ | $H_V$ | Ref. |
|---|---|---|---|---|---|---|---|---|---|---|---|---|---|---|
|  | $B_V$ | $B_R$ | $B_H$ | $G_V$ | $G_R$ | $G_H$ |  |  |  |  |  |  |  |  |
| TaP | 198.27 | 194.70 | 196.49 | 117.07 | 109.04 | 113.06 | 284.58 | 1.07 | 1.02 | 0.58 | 1.78 | 0.26 | 18.07 | This |
|  | - | - | 219 | - | - | 66 | 180 | - | - | - | - | 0.36 | - | [35]<sup>Theo.</sup> |
| TaAs | 169.63 | 166.29 | 167.96 | 96.70 | 87.06 | 91.88 | 233.13 | 1.11 | 1.02 | 0.55 | 1.93 | 0.27 | 14.07 | This |
|  | - | - | 189 | - | - | 54 | 147 | - | - | - | - | 0.37 | - | [35]<sup>Theo.</sup> |
|  | - | - | 197 | - | - | 53 | 145 | - | - | - | - | 0.38 | - | [34]<sup>Theo.</sup> |

## *3.3 Elastic anisotropy*

Elastic anisotropy explains directional dependence of physical properties of a compound. It is essential to study the anisotropic elastic properties of a material because a number of physical properties such as formation of micro-cracks in solids, motion of cracks, development of plastic deformations in crystals etc. are related to it. For instance, the degree of anisotropy in the bonding strength for atoms located in different planes is measured by the shear anisotropic factors. A proper explanation of these properties has significant implications in crystal physics and applied engineering sciences. Therefore, it is important to calculate elastic anisotropy factors of TaP and TaAs in details to understand their durability and possible applications under different external environments.

The degree of anisotropy in the bonding between atoms in different planes can be measured by shear anisotropic factors. The shear anisotropy for a tetragonal crystal can be quantified by three different factors [64, 84]:

The shear anisotropic factor for {100} shear planes between the ⟨011⟩ and ⟨010⟩ directions is,

$$A_1 = \frac{4C_{44}}{C_{11} + C_{33} - 2C_{13}} \quad (11)$$

The shear anisotropic factor for the {010} shear plane between ⟨101⟩ and ⟨001⟩ directions is,

$$A_2 = \frac{4C_{55}}{C_{22} + C_{33} - 2C_{23}} \quad (12)$$

and the shear anisotropic factor for the {001} shear planes between ⟨110⟩ and ⟨010⟩ directions is,

$$A_3 = \frac{4C_{66}}{C_{11} + C_{22} - 2C_{12}} \quad (13)$$

The calculated shear anisotropic factors of TaP and TaAs are enlisted in Table 4. For a crystal with isotropy in bonding between atoms in different planes have unit values ($A_1 = A_2 = A_3 = 1$) and any other value (lesser or greater than unity) implies degree of anisotropy possessed by the crystal. The estimated values of $A_1$, $A_2$ and $A_3$ predict that both the compounds are moderately anisotropic. TaP shows maximum anisotropy for $A_1$ and for TaAs, $A_1 = A_2$. Besides, there are several other useful anisotropy indices.

The universal log-Euclidean index is defined as [85, 86],

$$A^L = \sqrt{\left[\ln\left(\frac{B^V}{B^R}\right)\right]^2 + 5\left[\ln\left(\frac{C_{44}^V}{C_{44}^R}\right)\right]^2} \quad (14)$$

where, the Voigt and Reuss values of $C_{44}$ are obtained from [87]:

$$C_{44}^V = C_{44}^R + \frac{3}{5}\frac{(C_{11} - C_{12} - 2C_{44})^2}{3(C_{11} - C_{12}) + 4C_{44}} \quad (15)$$

and

$$C_{44}^R = \frac{5}{3}\frac{C_{44}(C_{11} - C_{12})}{3(C_{11} - C_{12}) + 4C_{44}} \quad (16)$$

The expression for $A^L$ is valid for all crystal symmetries. $A^L$ is closely related to the universal anisotropy index $A^U$. For a perfectly anisotropic crystal, $A^L=0$. The values of $A^L$ for TaP and TaAs indicate low level of anisotropy. The values of $A^L$ range between 0 to 10.26, and for almost 90% of the solids $A^L < 1$. It has been argued that $A^L$ is also an indicator regarding the layered/lamellar type of configuration [87]. Compounds with higher and lower $A^L$ values show strong layered and non layered structure, respectively. From the comparatively lower value of $A^L$, we can predict that our compounds do not exhibit layered type of structural configuration. The universal anisotropy index $A^U$, equivalent Zener anisotropy measure $A^{eq}$, anisotropy in compressibility $A^B$ and anisotropy in shear $A^G$ (or $A^C$) for the crystal with any symmetry are calculated using following standard equations [85, 88-90]:

$$A^U = 5\frac{G_V}{G_R} + \frac{B_V}{B_R} - 6 \geq 0 \qquad (17)$$

$$A^{eq} = \left(1 + \frac{5}{12}A^U\right) + \sqrt{\left(1 + \frac{5}{12}A^U\right)^2 - 1} \qquad (18)$$

$$A^B = \frac{B_V - B_R}{B_V + B_R} \qquad (19)$$

$$A^G = \frac{G^V - G^R}{2G^H} \qquad (20)$$

Universal anisotropy factor has become an attractive anisotropy index because of its simplicity compared to the plurality of anisotropy factors defined for specific planes in crystals. Ranganathan and Ostoja-Starzewski [88] introduced the concept of universal anisotropy index $A^U$ which provides a singular measure of anisotropy irrespective of the crystal symmetry. $A^U$ introduced the influence of the bulk to the anisotropy of a solid for the very first time unlike all other anisotropy measures. From Eqn. 14 it is seen that a larger fractional difference between the Voigt and Reuss estimated bulk or shear modulus would indicate a stronger degree of crystal anisotropy. From the values of $G_V/G_R$ and $B_V/B_R$ for TaP and TaAs, we can say that $G_V/G_R$ has more influence on $A^U$ than that due to $B_V/B_R$. For an isotropic crystal, $A^U=0$. While any deviation from this value, which must be positive, suggests presence and level of anisotropy. $A^U$ for TaP and TaAs are 0.39 and 0.57, respectively. These derivate from zero and both the compounds possess anisotropy in elastic/mechanical properties.

For an isotropic crystal, $A^{eq} = 1$. The calculated values of $A^{eq}$ for TaP and TaAs are 1.75 and 1.97, respectively predicting that both the compounds are anisotropic. For an isotropic crystal, $A^G = A^B = 0$. While any deviation greater than zero represents the degree of anisotropy. For both TaP and TaAs, $A^G$ is higher than $A^B$ (Table 4), which indicates that anisotropy in shear is larger than the anisotropy in compressibility. Compared to all other anisotropy measures; for both the compounds, $A^B$ predicts least anisotropy.

The linear compressibility of a tetragonal compound along $a$ and $c$ axis ($\beta_a$ and $\beta_c$) are calculated from [91]:

$$\beta_a = \frac{C_{33} - C_{13}}{D} \quad \text{and} \quad \beta_c = \frac{C_{11} + C_{12} - 2C_{13}}{D} \tag{21}$$

with $D = (C_{11} + C_{12})C_{33} - 2(C_{13})^2$

From the calculated values (Table 4) we can say that for both the compounds, compressibility along $a$ axis is lower than that along $c$ axis and also TaAs is more compressible than TaP. Which are also in good agreement with the calculated elastic constants of TaP and TaAs.

**Table 4**

Shear anisotropic factors ($A_1$, $A_2$ and $A_3$), universal log-Euclidean index $A^L$, the universal anisotropy index $A^U$, equivalent Zener anisotropy measure $A^{eq}$, anisotropy in shear $A^G$ (or $A^C$) and anisotropy in compressibility $A^B$, linear compressibilities ($\beta_a$ and $\beta_c$) (TPa$^{-1}$) and their ratio ($\beta_c/\beta_a$) for TaX (X = P, As) at P = 0 GPa and T = 0 K.

| Compounds | $A_1$ | $A_2$ | $A_3$ | $A^L$ | $A^U$ | $A^{eq}$ | $A^G$ | $A^B$ | Layered | $\beta_a$ | $\beta_c$ | $\beta_c/\beta_a$ | Ref. |
|---|---|---|---|---|---|---|---|---|---|---|---|---|---|
| TaP | 1.43 | 1.19 | 1.26 | 0.09 | 0.39 | 1.75 | 0.04 | 0.009 | No | 1.42 | 2.29 | 1.61 | This |
| TaAs | 1.18 | 1.18 | 1.31 | 0.12 | 0.57 | 1.97 | 0.05 | 0.010 | No | 1.64 | 2.73 | 1.66 | This |

The uniaxial bulk modulus along $a$, $b$ and $c$ axis and anisotropies of the bulk modulus are calculated by the following equations [84]:

$$B_a = a\frac{dP}{da} = \frac{\Lambda}{1 + \alpha + \beta} \quad ; \quad B_b = a\frac{dP}{db} = \frac{B_a}{\alpha} \quad ; \quad B_c = c\frac{dP}{dc} = \frac{B_a}{\beta} \tag{22}$$

and

$$A_{B_a} = \frac{B_a}{B_b} = \alpha \quad ; \quad A_{B_c} = \frac{B_c}{B_b} = \frac{\alpha}{\beta} \tag{23}$$

where, $\Lambda = C_{11} + 2C_{12}\alpha + C_{22}\alpha^2 + 2C_{13}\beta + C_{33}\beta^2 + 2C_{33}\alpha\beta$ and for tetragonal crystals, $\alpha = 1$ and $\beta = \frac{C_{11}+C_{12}-2C_{13}}{C_{33}-C_{13}}$. $A_{B_a}$ and $A_{B_c}$ represent anisotropies of bulk modulus along the *a* axis and *c* axis with respect to *b* axis, respectively. The calculated values are listed in Table 5. For both the compounds, $A_{B_a} = 1$ and $A_{B_b} \neq 1$, which indicates anisotropy in axial bulk modulus. Bulk modulus along *c* axis is smaller than those along *a* and *b* axis. These values are different from the isotropic bulk modulus and are much larger. This arises from the fact that the pressure in a state of uniaxial strain for a given crystal density generally differs from the pressure in a state of hydrostatic stress at the same density of the solid [92]. These findings are also in good accord with the result we got for linear compressibility.

**Table 5**

Anisotropies of bulk modulus along different axes of TaP and TaAs compounds.

| Compounds | $B_a$ | $B_b$ | $B_c$ | $A_{B_a}$ | $A_{B_c}$ | Ref. |
|---|---|---|---|---|---|---|
| TaP | 927.03 | 927.03 | 475.4 | 1 | 0.51 | This |
| TaAs | 722.73 | 722.73 | 44.23 | 1 | 0.60 | This |

## *3.4 Acoustic behavior and its anisotropy*

Sound velocity is an important property of a material related to its thermal and electrical conductivity. Acoustic behavior of a material has become a matter of notable interest in physics, materials science, seismology, geology, musical instrument designing and medical science. The transverse and longitudinal velocity of sound traversing through a crystalline solid with bulk modulus *B* and shear modulus *G* can be written as [93]:

$$v_t = \sqrt{\frac{G}{\rho}} \quad \text{and} \quad v_l = \sqrt{\frac{B + 4G/3}{\rho}} \tag{24}$$

where $\rho$ refers to the mass-density of the solid.

The average sound velocity $v_a$ can be evaluated from the transverse and longitudinal sound velocities using [93]:

$$v_a = \left[\frac{1}{3}\left(\frac{2}{v_t^3} + \frac{1}{v_l^3}\right)\right]^{-\frac{1}{3}} \tag{25}$$

A material having same or different acoustic impedance with respect to the surrounding medium is an important factor of transducer design, noise reduction in aircraft engine, industrial factories and many underwater acoustic applications. When sound is transmitting from one material to another, the amount of transmitted and reflected sound energy at the interface depends on their difference in acoustic impedance. Therefore, if the two impedances are about equal, most of the sound gets transmitted, but if the impedances differ greatly, most of it is reflected. The acoustic impedance of a material can be defined as [94]:

$$Z = \sqrt{\rho G} \qquad (26)$$

This equation implies that a material with high density and high shear modulus has high acoustic impedance.

The intensity of sound radiation is another important parameter for designing sound boards. The intensity, $I$, is proportional to the surface velocity for a given driving function, this scales with modulus of rigidity and density as [95]:

$$I \approx \sqrt{G/\rho^3} \qquad (27)$$

A high value of $\sqrt{G/\rho^3}$, called the *radiation factor*, is used by instrument makers to select materials for suitably designed sound boards.

The calculated sound velocities, acoustic impedance and radiation factor for TaP and TaAs are listed in Table 6.

**Table 6**

Density $\rho$ (g/cm$^3$), transverse velocity $v_t$ (ms$^{-1}$), longitudinal velocity $v_l$ (ms$^{-1}$), average elastic wave velocity $v_a$ (ms$^{-1}$), acoustic impedance $Z$ (Rayl) and radiation factor $\sqrt{G/\rho^3}$ (m$^4$/kg.s) of TaP and TaAs compounds.

| Compounds | $\rho$ | $v_t$ | $v_l$ | $v_a$ | $Z$ (×10$^6$) | $\sqrt{G/\rho^3}$ | Ref. |
|---|---|---|---|---|---|---|---|
| TaP | 9.86 | 3386.22 | 5934.37 | 3732.31 | 33.39 | 0.34 | This |
|  | 10.02 | - | - | - | - | - | [96]$^{Exp.}$ |
| TaAs | 10.81 | 2915.40 | 5183.65 | 3217.71 | 31.52 | 0.27 | This |
|  | 11.65 | - | - | - | - | - | [96]$^{Exp.}$ |

The propagation velocity of sound (longitudinal and transverse) waves in a solid is independent of its frequency and the dimension of the material and depends only on the nature of the material. Every atom in a solid has three modes of vibrations, one longitudinal and two transverse modes. Elastically anisotropic solids have anisotropic sound velocities. Also for anisotropic solids, sound velocities can only propagate in pure longitudinal and transverse mode along certain crystallographic directions. For tetragonal symmetry, the pure transverse and longitudinal modes can be found for [010], [100], [001], and [110] directions. For tetragonal crystal, the acoustic velocities along these principle directions can be expressed as [97]:

[010] = [100]:

$[100]v_l = [010]v_l = \sqrt{C_{11}/\rho}; [001]v_{t1} = \sqrt{C_{44}/\rho}; [010]v_{t2} = \sqrt{C_{66}/\rho}$

[001]:

$[001]v_l = \sqrt{C_{33}/\rho}; [010]v_{t2} = [100]v_{t1} = \sqrt{C_{66}/\rho}$ (28)

[110]:

$[110]v_l = \sqrt{(C_{11} + C_{12} + 2C_{66})/2\rho}; [001]v_{t1} = \sqrt{C_{44}/\rho}; [1\bar{1}0]v_{t2} = \sqrt{(C_{11} - C_{12})/2\rho}$

where $v_{t1}$ and $v_{t2}$ are the first transverse mode and the second transverse mode, respectively.

Both longitudinal and transverse sound velocities are correlated to elastic constants and crystal density, the compound with low density and high elastic constants will have large sound velocities.

**Table 7**

Anisotropic sound velocities (ms$^{-1}$) of TaP and TaAs along different crystallographic directions.

| Propagation directions | | TaP | TaAs |
|---|---|---|---|
| [100] | $[100]v_l$ | 5856.12 | 5081.57 |
| | $[001]v_{t1}$ | 3342.36 | 2838.22 |
| | $[010]v_{t2}$ | 4312.11 | 3928.12 |
| [001] | $[001]v_l$ | 5347.46 | 4656.67 |
| | $[100]v_{t1}$ | 4312.11 | 3928.12 |
| | $[010]v_{t2}$ | 4312.11 | 3928.12 |
| [110] | $[110]v_l$ | 6637.41 | 5931.03 |
| | $[1\bar{1}0]v_{t2}$ | 2972.06 | 2476.44 |
| | $[001]v_{t1}$ | 3342.36 | 2838.22 |

## 3.5 Thermal properties

### 3.5.1 Debye temperature

Debye temperature ($\Theta_D$) is one of the most prominent thermo-physical parameter of solids, closely related to large number of physical properties like, thermal conductivity, lattice vibration, interatomic bonding, melting temperature, coefficient of thermal expansion and phonon specific heat. Generally, a compound with stronger interatomic bonding strength, lower average atomic mass, higher melting temperature, greater hardness and higher mechanical wave velocity has larger Debye temperature. At low temperature, $\Theta_D$ calculated from elastic constants is the same as that calculated from specific heat, since at low temperatures the vibrational excitations arise solely from acoustic modes (lattice vibrations). The Debye temperature can be obtained from the average sound velocity using the following equation [93, 98]:

$$\Theta_D = \frac{h}{k_B}\left(\frac{3n}{4\pi V_0}\right)^{1/3} v_a \qquad (29)$$

where, $h$ is Planck's constant, $k_B$ is the Boltzmann's constant, $V_0$ is the volume of unit cell and $n$ is the number of atoms within the unit cell.

The calculated Debye temperature of TaP and TaAs are listed in Table 8. Debye temperature of TaP is much higher than TaAs. Thus it is expected that lattice thermal conductivity of TaP should be significantly higher than TaAs.

### 3.5.2 Melting temperature

In a theoretical search for new materials to be used at different temperatures, an interesting and important area of research is to specify the melting temperature. Compounds with higher melting temperature have lower thermal expansion and higher bonding energy. Fine *et al.* showed that the average elastic moduli ($C_{11}+C_{22}+C_{33}$)/3 and the melting temperature of a solid are interrelated. We have calculated melting temperature $T_m$ using following equation [99]:

$$T_m = 354K + (4.5K/GPa)\left(\frac{2C_{11} + C_{33}}{3}\right) \pm 300K \qquad (30)$$

The calculated melting temperatures of TaP and TaAs are listed in Table 8. The melting temperature of TaP and TaAs are 1791.35 K and 1544.90 K, respectively indicating that both are good candidate materials for high temperature applications. This also agrees with bulk modulus, Debye temperature, minimum thermal conductivity, strength of the bonds and hardness. The bonding energy of crystalline material proportionally related with its melting temperature $T_m$.

### 3.5.3 Thermal expansion and Heat capacity

The thermal expansion (TE) of a material is connected to many other physical properties, like, thermal conductivity, specific heat, temperature variation of the energy band gap and electron effective mass. Thermal expansion coefficient ($\alpha$) is also important for epitaxial growth of crystals, to reduce the harmful effects during its use in the electronic and spintronic devices. The thermal expansion coefficient of a material can be obtained using the following equation [94]:

$$\alpha = \frac{1.6 \times 10^{-3}}{G} \tag{31}$$

Thermal expansion is inversely related to melting temperature: $\alpha \approx 0.02/T_m$ [94, 99]. From Table 8, we can see that TaP with higher melting temperature has lower thermal expansion compared with TaAs.

Heat capacity is another important thermodynamic parameter. A compound with higher heat capacity has higher thermal conductivity and lower thermal diffusivity. The heat capacity per unit volume can be calculated from following equation [94]:

$$\rho C_P = \frac{3k_B}{\Omega} \tag{32}$$

where, $N = 1/\Omega$ is the number of atoms per unit volume.

**Table 8**

The Debye temperature $\Theta_D$(K), thermal expansion coefficient $\alpha$(K$^{-1}$), melting temperature $T_m$(K) and heat capacity per unit volume $\rho C_P$(J/m$^3$.K) of TaP and TaAs compounds.

| Compounds | $\Theta_D$ | $\alpha$ (×10$^{-5}$) | $T_m$ | $\rho C_P$ (×10$^{-6}$) | Ref. |
|---|---|---|---|---|---|
| TaP | 276.10 | 1.415 | 1791.35 | 2.55 | This |
|  | 408 | - | - | - | [35]$^{Theo.}$ |
| TaAs | 230.57 | 1.741 | 1544.90 | 2.32 | This |
|  | 341 | - | - | - | [35]$^{Theo.}$ |
|  | 338 | - | - | - | [34]$^{Theo.}$ |

### 3.5.4 Minimum thermal conductivity

At high temperatures above the Debye temperature, thermal conductivity of a compound approaches a minimum value known as minimum thermal conductivity ($k_{min}$). One striking feature of the minimum thermal conductivity is that it does not depend on the presence of defects (such as dislocations, individual vacancies and long-range strain fields associated with inclusions and dislocations) inside the crystal. This is largely because these defects affect phonon transport over length scales much larger than the interatomic spacing and at high temperatures the phonon mean free path becomes significantly smaller than this length scale. Based on the Debye model, Clarke deduced the following formula for calculating the minimum thermal conductivity $k_{min}$ of compounds at high temperatures [100]:

$$k_{min} = k_B v_a (V_{atomic})^{-2/3} \qquad (33)$$

In this equation, $k_B$ is the Boltzmann constant, $v_a$ is the average sound velocity and $V_{atomic}$ is the cell volume per atom [100].

The calculated values of minimum thermal conductivity for TaX (X = P, As) WSMs are enlisted in Table 9. $k_{min}$ of TaP is higher than TaAs. Compounds with higher sound velocity and Debye temperature have higher minimum thermal conductivity.

It is well known that heat is transmitted through solids in three different modes: by thermal vibrations of atoms, by the movement of free electrons in metals, and, if they are transparent, by radiation. Transmission by thermal vibrations involves the propagation of elastic waves. An elastically anisotropic material also has anisotropic minimum thermal conductivity. The anisotropy in minimum thermal conductivity depends on sound velocity in different crystallographic directions. The minimum thermal conductivities along different directions are calculated by using Cahill and Clarke model [101]:

$$k_{min} = \frac{k_B}{2.48} n^{2/3} (v_l + v_{t1} + v_{t2}) \qquad (34)$$

and $\qquad n = N/V$

where, $k_B$ is the Boltzmann constant, $n$ is the number of atoms per unit volume and $N$ is total number of atoms in the cell having a volume $V$.

The minimum thermal conductivity of TaX (X = P, Ta) along [100], [001] and [110] directions are summarized in Table 9. Minimum thermal conductivity of TaP is higher than that of TaAs. For both the compounds, the minimum thermal conductivities along different crystallographic directions are higher than the isotropic minimum thermal conductivity.

**Table 9**

The number of atom per mole of the compound $n$ (m$^{-3}$), minimum thermal conductivity (W/m.K) of TaP and TaAs compounds along different directions evaluated by the Cahill's method and Clarke method.

| Compound | $n(10^{28})$ | $[100]k_{min}$ | $[001]k_{min}$ | $[110]k_{min}$ | $k_{min}$ | Ref. |
|---|---|---|---|---|---|---|
| TaP | 6.167 | 1.176 | 1.216 | 1.127 | 0.822 | This |
| TaAs | 5.604 | 0.968 | 1.022 | 0.919 | 0.665 | This |

## *3.6 Electronic Properties*

### *3.6.1 Band structure*

In solid state physics, electronic band structure is one of the most important concepts that can explain many electrical, optical, and even some magnetic properties of crystals. The electronic band structure, as a function of energy ($E$-$E_F$), along different high symmetry directions ($M$-$\Gamma$-$X$-$M$-$R$-$X$-$\Gamma$-$R$) in the first BZ has been calculated at zero pressure and temperature. Figs. 2(a) and 2(b) show the band structure of TaP and TaAs, respectively. The horizontal broken lines indicate the Fermi level ($E_F$). The total number of bands for TaP and TaAs are 61 and 79, respectively. From Fig. 2(a and b), we can see that there is no band gap at the Fermi level, due to the overlap of valence band and conduction band. The extent of band overlap and crossings of the Fermi level are rather weak. This implies that Ta$X$ ($X$ = P, As) compounds are expected to exhibit semimetallic character. For both TaP and TaAs, the $M$-$\Gamma$ and $X$-$M$-$R$ bands do not contribute in charge transport whereas $\Gamma$–$X$ and $R$-$X$-$\Gamma$-$R$ bands contributes to electrical and thermal conduction. Moreover, energy bands around the Fermi level are mainly derived from the Ta 5$d$ states for both the compounds. This indicates that the Ta 5$d$ states dominate the conductivity of TaP and TaAs. On the other hand, bands below the Fermi level mainly originate from the Ta 5$d$ and P/As 3$p$ for TaP and TaAs, respectively. Bands running along $X$-$M$ and $M$-$R$ show significant electronic dispersion near the Fermi energy. Several parts of the $E(k)$ curves show linear dispersion. For example, there is a linear electronic dispersion near $X$ point (shown in blue) of TaP and TaAs, respectively. These linear dispersions are due to Dirac cones which prime signature features of topological semimetals.

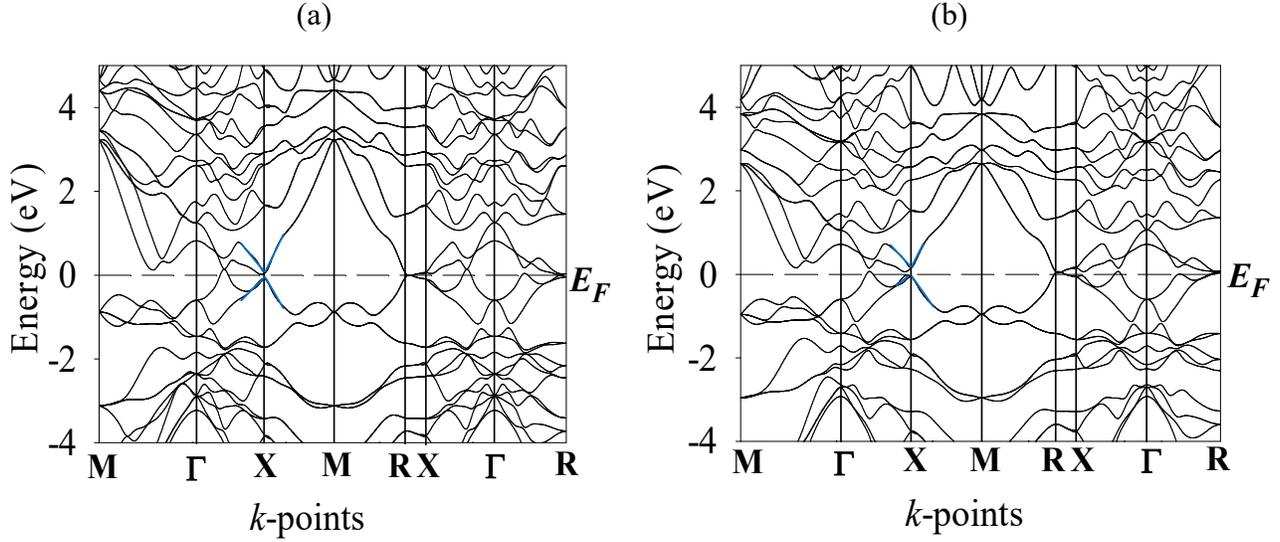

**Figure 2.** Electronic band structures of (a) TaP and (b) TaAs along several high symmetry directions in the Brillouin zone at P = 0 GP. Representative Dirac-cones are marked in blue.

### 3.6.2 Density of states (DOS)

The calculated total and partial density of states (TDOS and PDOS, respectively) of TaP and TaAs at zero pressure and temperature is shown in Figures 3(a) and 3(b), respectively. The vertical broken lines denote the Fermi level. The non-zero values of TDOS at the Fermi level indicate that both TaP and TaAs will exhibit metallic electrical conductivity. To understand the contribution of each atom in the TDOS of TaP and TaAs, we have also calculated the PDOS of Ta, P and As atoms. TDOS values of TaP and TaAs at the Fermi level are comparable and 1.77 and 1.70 states per eV per unit cell, respectively. Near the Fermi level, the main contribution comes from Ta $5d$ and P/As $3p$ states in TaP and TaAs. There is significant hybridization between these electronic states. Such hybridization near the Fermi energy is often indicative of formation of covalent bonding.

Electronic stability of a compound is associated with the position of Fermi level and the value of the TDOS at the Fermi energy, $N(E_F)$ [102, 103]. The electronic stability of a solid is related to the presence of a pseudogap or quasi-gap in the TDOSs around the Fermi level [104, 105]. This gap separates bonding states from nonbonding/antibonding electronic states. For both TaP and TaAs, Fermi levels lie almost on the pseudogap minima, particularly for TaAs (see Figs. 3). Two mechanisms [106, 107] are usually mentioned in the literature responsible for the origin of the pseudogap or the quasi-gap. One has an ionic origin and the other is due to hybridization among atomic orbitals. The pseudogap around the Fermi level also indicates towards the presence of the directional bonding [108] which facilates the formation of covalent bonding and enhances the mechanical strength of material. To be specific, around the Fermi level, the strong bonding hybridization is originated mainly from the Ta $5d$ states with P/As $3p$ states, which thus forms the directional covalent bonding between Ta-P and Ta-As atoms in TaP and TaAs, respectively. This agrees with the electronic charge density mapping and Mulliken bond population analysis results (presented in Section 3.6.3 and 3.7). If the Fermi level lies exactly at the pseudogap, it refers to an ordered compound with high melting point. If Fermi level lie to the right of the pseudogap, i.e. antibonding region, it refers to an unstable state and the system will exhibit tendency to be in the disordered or glassy state [109].

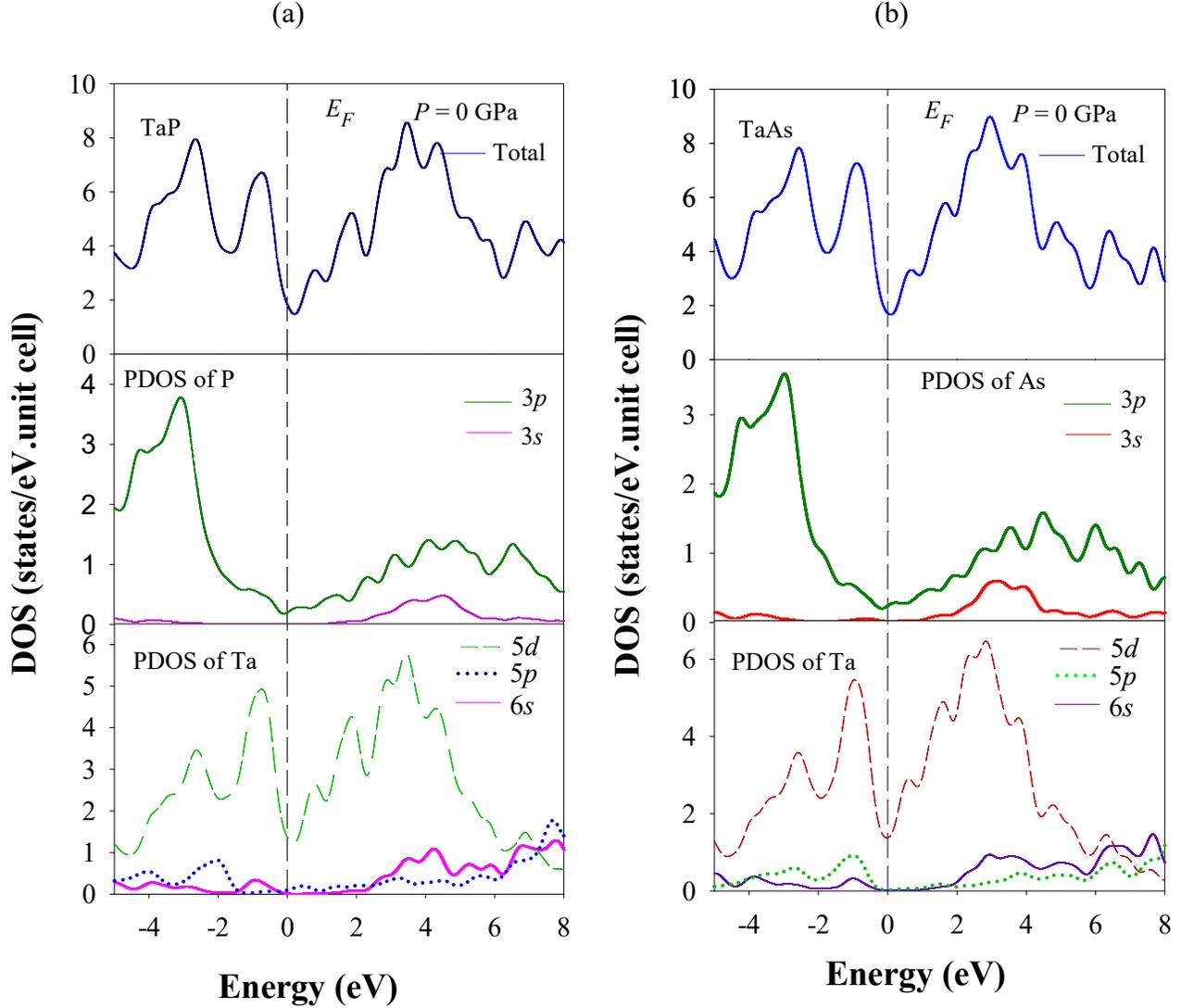

**Figure 3.** Total and partial electronic density of states (PDOSs and TDOSs) of (a) TaP and (b) TaAs at P = 0 GPa

The electron-electron interaction parameter can be estimated using the following relation [110]:

$$\mu^* = \frac{0.26 N(E_F)}{1 + N(E_F)} \tag{35}$$

The total density of states at the Fermi level for TaP and TaAs are 1.77 and 1.70 states/eV.unit cell, respectively. The electron-electron interaction parameters of TaP and TaAs are found to be 0.17 and 0.16, respectively. This repulsive Coulomb pseudopotential reduces the transition temperature, $T_c$ of superconducting compounds [109-111].

### *3.6.3 Electronic charge density distribution*

To attain further insights into the bonding nature of Ta$X$ ($X$ = P, As), the charge distribution around the atoms within the crystal has been investigated. We have studied electronic charge density ($e/Å^3$) distribution within different crystal planes. Figure 4 shows the electronic charge density distribution of TaP and TaAs in the (100) and (010) planes. The color scales on the right hand side of each compound's charge density maps illustrate the total electron density. The red color indicates high charge (electron) density and blue color indicates low charge (electron) density for TaP and for TaAs, respectively. The charge density distribution map shows clear signatures of covalent bonding between Ta-Ta, Ta-P and P-P atoms and Ta-Ta, Ta-As and As-As in TaP and TaAs compounds respectively. These findings agree with the Mulliken bond population analysis (Section 3.7). For both the planes, from the charge density maps of TaP and TaAs, we can see that P and As atoms have high electron density compared to Ta atoms. Ta atoms are electron deficient in both TaP and TaAs and have nearly spherical charge distribution. These point toward some ionic contribution to the overall bonding. The charge distribution around P and As are non-spherical, which suggests directionality and covalent bonding between Ta and P atoms and Ta and As atoms, respectively. This feature is consistent with the DOS curves (Fig. 3a), which shows strong hybridization between Ta-5$d$ and P-3$p$ in TaP and Ta-5$d$ and As-3$p$ in TaAs. Charge density distribution of TaP and TaAs on different planes shows both direction and plane dependency. Thus we can say that, both TaP and TaAs have anisotropic charge density distribution.

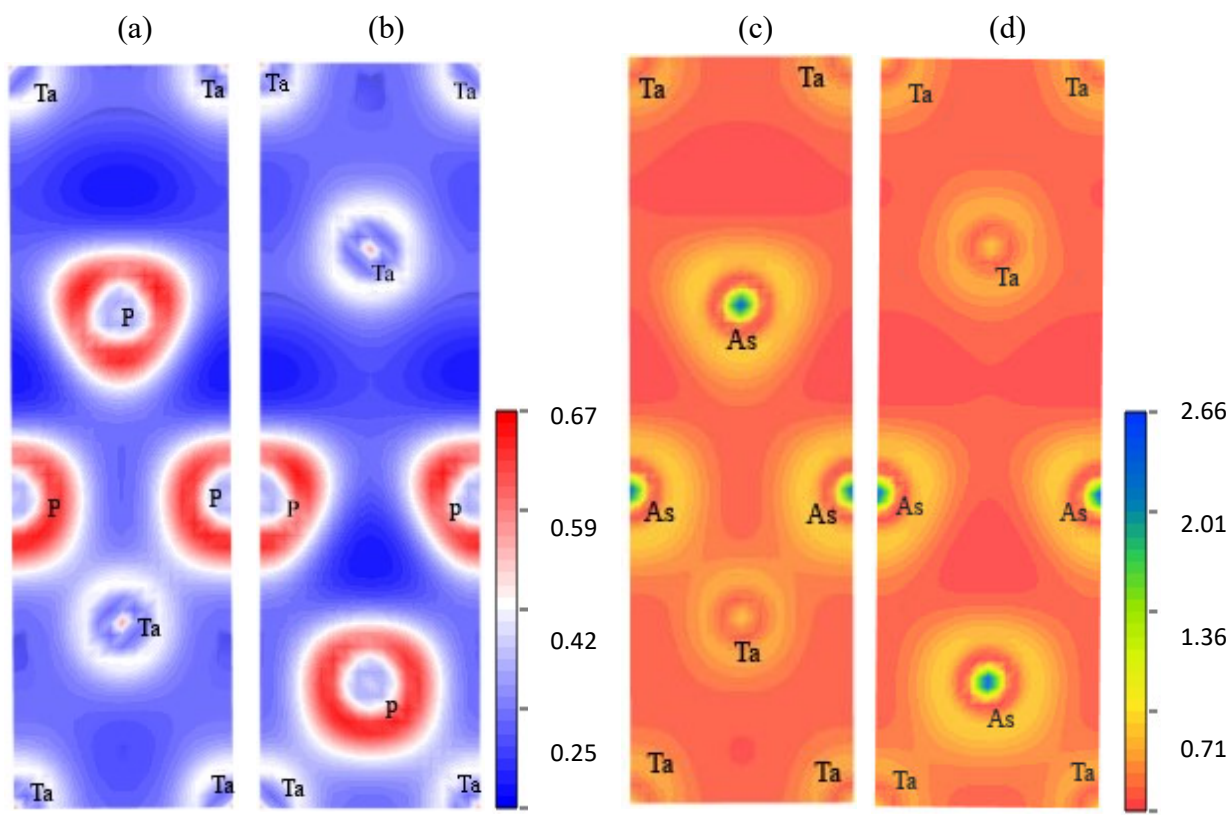

**Figure 4.** The electronic charge density distribution map for TaP (a and b) and TaAs (c and d) in the (100) and (010) planes, respectively.

## *3.7 Bond population analysis*

The Mulliken bond populations [112] are studied to get understanding of the bonding nature (ionic, covalent and metallic) in a compound and effective valence of atoms in the molecules present in the compound in greater depth. The Mulliken charge is related to the vibrational properties of the molecule and quantifies how the electronic structure changes under atomic displacement. It also affects dipole moment, polarizability, and charge mobility in reactions, electronic structure, and other related properties of molecular systems. The results of this analysis are tabulated in Table 10. It is found that the total charge for P and Ta atoms are larger than those for Ta and As atoms in TaP and TaAs, respectively. The atomic charge of Ta and P in TaP are 0.07 and -0.07 electron, respectively. Both are deviated from the normal value expected for purely ionic state (Ta: +5 and P: +5, +3, -3). This deviation reflects that covalent bonds are present between Ta and P with ionic contribution. For TaP, electrons are transferred from Ta to P. Therefore, we can predict some ionic contribution in the bonding between Ta and P in TaP.

From orbital charge values, we can say that these electrons mainly come from $5d$ orbital of Ta. Similarly the atomic charge of Ta and As in TaAs are -0.52 and 0.52 electron, respectively. Both of these values are deviated from the normal value expected for purely ionic state (Ta: +5 and As: +5, +3, -3). This reflects the presence of covalent bond between Ta and As in TaAs. For TaAs, electrons are transferred from As to Ta. The contribution of ionic bonding is higher in TaAs. From orbital charge values, we can assume that these electrons mainly come from the $3p$ orbital of As.

To comprehend the degree of covalency and/or iconicity, the effective valence of TaP and TaAs are calculated. Effective valence is defined as the difference between the formal ionic charge and the Mulliken charge on the cation species [55]. The zero value of effective valence indicates perfect ionic bond, while values greater that zero indicate increasing level of covalency. The effective valences for Ta in TaP and for As in TaAs are +4.93 and +4.48, respectively. This implies that both ionic and covalent bonds are present in TaP and TaAs and covalent bondings dominate in these compounds.

Since early days, it was recognized that the Mulliken bond population analysis has strong basis set dependency and because of this, sometimes it gives results in contradiction to chemical intuition. On the other hand, Hirshfeld population analysis (HPA) gives more meaningful result because it has practically no basis set dependence. Keeping this point in mind, we have determined Hirshfeld charge of TaP and TaAs using the HPA. For TaAs, HPA shows opposite result compared to Mulliken charge. For TaP, Hirshfeld analysis shows atomic charge of Ta and P is + 0.11 and -0.11 electronic charge, respectively. For TaAs, Hirshfeld analysis shows atomic charge of Ta and As is + 0.08 and -0.08 electronic charge, respectively. Hirshfeld charge predicts that electrons are transferred from Ta to P and As atoms in TaP and TaAs, respectively. At the same time, it should be noted that both the approaches predict an admixture of covalent and ionic bondings between Ta and P/As atoms. We have also calculated effective valences of TaP and TaAs using the Hirshfeld charge, which are +4.89 and +4.92 electron, respectively. Therefore, HPA predicts that TaP has lower level of covalency compared to TaAs, which is different from the result we get from the Mulliken charge analysis.

**Table 10**

Charge spilling parameter (%), orbital charges (electron), atomic Mulliken charges (electron), effective valence (electron) and Hirshfeld charge (electron) in TaP and TaAs.

| Compounds | Species | Charge spilling | s | p | d | Total | Mulliken charge | Effective valence | Hirshfeld charge | Effective valence |
|---|---|---|---|---|---|---|---|---|---|---|
| TaP | Ta | 0.86 | 0.47 | 0.43 | 4.03 | 4.93 | 0.07 | +4.93 | 0.11 | +4.89 |
|  | P |  | 1.60 | 3.47 | 0.00 | 5.07 | -0.07 |  | -0.11 |  |
| TaAs | Ta | 0.65 | 0.68 | 0.83 | 4.01 | 5.52 | -0.52 |  | 0.08 |  |
|  | As |  | 1.05 | 3.43 | 0.00 | 4.48 | 0.52 | +4.48 | -0.08 | +4.92 |

## *3.8 Theoretical bond hardness*

The study of the behavior of a material under varying load is an important part to understand its application, especially using as an abrasive resistant phase and radiation tolerant material [113]. A compound with higher bond density or electronic density, shorter bond length, and greater degree of covalent bonding is harder. In general, a compound with larger bulk modulus and shear modulus indicates higher hardness. Though the bulk modulus or shear modulus gives some information regarding hardness, there is no direct, one to one correspondence between hardness and bulk/shear modulus [114]. There are two types of hardnesses: intrinsic and extrinsic. Generally the hardness of a perfect crystal is considered as intrinsic and that of nanocrystalline and polycrystalline as extrinsic. The hardness of compound can be obtained using the following equations [115, 116]:

$$H_V = \left[\prod^{\mu}\left\{740(P^{\mu} - P^{\mu'})(v_b^{\mu})^{-5/3}\right\}^{n^{\mu}}\right]^{1/\Sigma n^{\mu}} \tag{36}$$

where $P^{\mu}$ is the Mulliken population of the $\mu$-type bond, $P^{\mu'} = n_{free}/V$ is the metallic population (with $n_{free}$= number of free electrons), $n^{\mu}$ is the number of $\mu$-type bond, and $v_b^{\mu}$ is the bond volume of $\mu$-type bond. The constant 740 is a proportional coefficient obtained from the hardness of diamond.

The calculated bond length, overlap population and the theoretical hardness of TaP and TaAs are given in Table 11. The hardness of covalent crystal depends on the sum of resistance of each bond per unit area to the indenter [116, 117]. The Mulliken bond populations define the degree of overlap degree of the electron clouds forming bondings between atoms in the crystal. The overlap population of electrons between atoms is a measurement of the strength of the covalent bond between atoms and the strength of the bond per the unit volume. The positive (+) and negative (-) values of overlap population indicate the presence of bonding-type and anti-bonding-type interactions between the atoms, respectively [118]. The overlap population close to zero indicates that there is no significant interaction between the electronic populations of the two bonding atoms. The calculated values indicate that bonding type interaction dominates in TaP. Whereas both bonding-type and anti-bonding-type interactions are present in TaAs. It is also clear that as far as bonding strengths are concerned, TaP is stronger than TaAs.

It is instructive to note that, from the effective valence concept, sometimes it is difficult to put a figure on the level of ionicity and covalency. Overlap population qualitatively reflect the trend of bond iconicity/covalency. A high positive overlap population indicates a low degree of iconicity or high degree of covalency (e.g., sharing of electrons via overlapping atomic orbitals between atoms) of the chemical bonding. From Table 10, we can say that covalent nature present in TaP is stronger than in TaAs.

Metallic bonding is soft in nature and has negligible contribution to the hardness of a material. Metallic population for both TaP and TaAs is very low. The calculated values of theoretical hardness of TaP and TaAs are shown in Table 10. The hardness of TaP is significantly higher than TaAs.

**Table 10**

The calculated Mulliken bond overlap population of $\mu$-type bond $P^\mu$, bond length $d^\mu$ (Å), metallic population $P^{\mu'}$, total number of bond $N^\mu$, bond volume $v_b^\mu$ (Å$^3$), hardness of $\mu$-type bond $H_v^\mu$ (GPa) and Vickers hardness of the compound, $H_v$ (GPa) of TaP and TaAs.

| Compounds | Bond | $P^\mu$ | $d^\mu$ | $P^{\mu'}$ | $N^\mu$ | $v_b^\mu$ | $H_v^\mu$ | $H_v$ |
|---|---|---|---|---|---|---|---|---|
| TaP | P-Ta | 0.62 | 2.55 | 0.061 | 8 | 16.129 | 4.014 | 15.976 |
| | P-Ta | 1.76 | 2.56 | 0.061 | | 16.320 | 11.962 | |
| TaAs | As-Ta | -0.14 | 2.63 | 0.061 | 8 | 17.641 | -1.243 | 4.407 |
| | As-Ta | 1.01 | 2.65 | 0.061 | | 18.047 | 5.650 | |

## *3.9 Optical Properties*

Study of optical properties of compounds has become a crucial part in materials science. It is also important to explore possible applications in optoelectronic and photovoltaic device sectors. Optical properties of a material describe mainly the interaction of electromagnetic wave with the charge carrier in materials. The presence of a medium in electric and magnetic fields may lead to electric dipoles and magnetic moments, polarization charges, and induced current. Clearly the electric and magnetic fields will not be uniform within the material but fluctuate from point to point reflecting the periodicity of the atomic lattice. Since both TaP and TaAs are elastically anisotropic, optical parameters may also show direction dependence. So we have investigated optical properties of TaP and TaAs for two different polarization directions [100] and [001] of the incident electric field. The electronic band structures properties TaP and TaAs show that both the compounds are metallic in nature. For metallic material, the Drude damping correction is required [119, 120]. The calculation of the optical constants of TaP and TaAs have done using a screened plasma energy of 10 eV and a Drude damping of 1 eV as prescribed in the CASTEP.

The macroscopic electronic response of a material can fully be described by the complex dielectric function,

$$\varepsilon(\omega) = \varepsilon_1(\omega) + i\varepsilon_2(\omega) \qquad (37)$$

where $\varepsilon_1(\omega)$ and $\varepsilon_2(\omega)$ are real and imaginary parts of dielectric function $\varepsilon(\omega)$. In the condensed matter system, there are two contributions to $\varepsilon(\omega)$, namely intra-band and inter-band transitions. Intra-band transitions play an important role at low energy and the inter-band term depends strongly upon the details of the electronic band structure [121]. For both TaP and TaAs, the large negative values of $\varepsilon_1$ at low energies indicate that the crystal has a Drude-like behavior, whereas at higher energies the inter-band transitions take place giving rise to optical features. On the other hand, the large positive values of $\varepsilon_2$ at low energies indicate that both the compounds have high absorption for the both polarizations. From complex dielectric function $\varepsilon(\omega)$ of a material, one can calculate other energy (frequency) dependent optical functions, such as the refractive index $n(\omega)$, extinction coefficient $k(\omega)$, optical reflectivity $R(\omega)$, absorption coefficient $\alpha(\omega)$, energy-loss function $L(\omega)$, and complex optical conductivity $\sigma(\omega)$. For a good metallic system, the dielectric contribution become less important compared to the conductive contribution at low frequencies. Figure 5(a) and 6(a) show the real and imaginary part of $\varepsilon(\omega)$ for photon energy up to 30 eV of TaP and TaAs, respectively.

The refractive index is a complex parameter, expressed as $N(\omega) = n(\omega) + ik(\omega)$, where the imaginary part $k(\omega)$ is known as extinction coefficient. The real part of the refractive index illustrates the phase velocity of the electromagnetic wave inside the sample, while the extinction coefficient (imaginary part) spectrum explains the amount of attenuation of the incident electromagnetic radiations when traveling through the material. The frequency dependence of refractive index (real and imaginary) of TaP and TaAs for both polarization directions are shown in Figure 5(b) and 5(b), respectively. For both TaP and TaAs, refractive index is large at low energy and decreases with increasing energy.

The conduction of free charge carriers over a defined range of the photon energies are explained by optical conductivity. This is a dynamic response of mobile charge carriers which includes the photon generated electron hole pairs in semiconductors. Figure 5(c) and 6 (c) display the calculated frequency-dependent optical conductivity $\sigma(\omega)$ of TaP and TaAs, respectively. For both TaP and TaAs, photoconductivity starts with zero photon energy, which indicates that the materials has no band gap agreeing with the band structure (Fig. 2) and TDOS calculations (Fig. 3). The photoconductivity of both the compounds increases with photon energy, reaches to maximum, decreases gradually with further increase in energy and tends to zero at around 20 eV. Both TaP and TaAs have higher conductivity along [100] direction compared to [001] direction, especially at low energies, indicating a small anisotropy in optical properties. The optical conductivity of a material can be related to its imaginary part of the dielectric function. This is obvious from qualitative agreement in the response spectra of $\sigma(\omega)$ and $\varepsilon_2(\omega)$.

Figure 5(d) and 6(d) show reflectivity spectrum of TaP and TaAs, respectively. For both the compounds, the reflectivity spectra start from zero frequency for both field polarizations. Reflectivities for both TaP and TaAs are somewhat higher for [100] polarization of the electric field. TaP shows high reflectivity compared to TaAs. $R(\omega)$ of both the compounds show almost nonselective behavior over this broad energy range.

The absorption coefficient is an important parameter to understand a material's electrical nature, whether it is metallic, semiconducting or insulating. It also helps us to understand the optimum solar energy conversion efficiency of a material. Figure 5(e) and 6(e) show the absorption spectra of TaP and TaAs, respectively. As we can see from the figures that for both TaP and TaAs, optical absorption starts from zero photon energy. This reconfirms the absence of optical band gap and semimetallic nature. The optical absorption starts due to the free electrons within the conduction band. The absorption coefficient of TaP is quite high in the region from ~5.9 to 19.4 eV, peaking around 8.04 eV. The absorption coefficient of TaAs is high in the spectral region from ~ 3.6 to 15 eV, peaking around 7.4 eV. For both the compounds, $\alpha(\omega)$ is slightly higher along the [100] polarization direction compared to [001]. $\alpha(\omega)$ decreases sharply at ~ 23.8 eV and ~23.5 eV for TaP and TaAs, respectively, in agreement with the position of the loss peak.

Figure 5(f) and 6(f) show the frequency dependent electron energy loss function $L(\omega)$ of TaP and TaAs, respectively. Loss function is an essential optical parameter describing the energy loss of a first electron traveling in a material. Loss function, absorption and reflection characteristics of a material are interrelated. The highest peak of loss spectrum represent the plasma resonance due to collective charge excitation and the corresponding frequency is known as the plasma frequency $\omega_P$ of that material. This appears at $\varepsilon_2 < 1$ and $\varepsilon_1 = 0$ [94, 122]. The energy at which $L(\omega)$ is maximum, is known as the plasmon energy. Study of energy loss function is useful for understanding the screened charge excitation spectra, particularly the collective excitations produced by a swift electron traversing a solid. The loss spectrum also arises due to photon absorption with appropriate energy which can give rise to the same excitations of collective charge oscillations called the plasmons. The peak in $L(\omega)$ spectra appears at a particular incident light frequency (energy), known as the bulk screened plasma frequency. It is observed that for TaP, the peaks of $L(\omega)$ for [100] and [001]polarizations are located at 24.73 eV and 24.38 eV, respectively. For TaAs, the peaks of $L(\omega)$ for [100] and [001]polarizations are located at 24.92 eV and 24.11 eV, respectively. Sharp loss peaks represent the abrupt reduction in reflectivity and absorption coefficient of TaP and TaAs (see Figs. 5(d) & (e) and 6(d) & (e), respectively).

Above this resonance energy, TaP and TaAs are expected to be transparent to incident photons and will change their response from metallic to dielectric like.

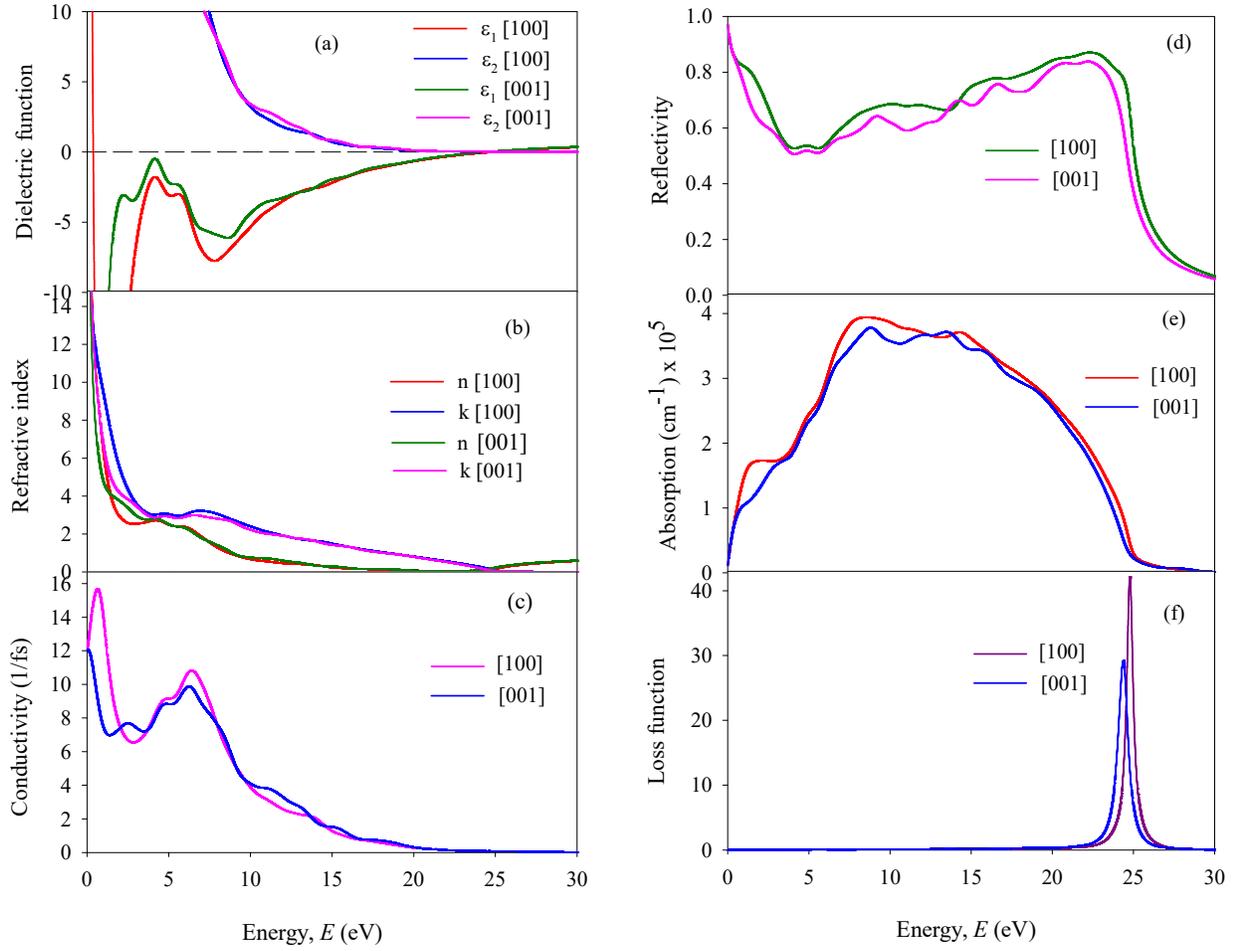

**Figure 5:** (a) $\varepsilon_1(\omega)$ and $\varepsilon_2(\omega)$, (b) $n(\omega)$ and $k(\omega)$, (c) $\sigma(\omega)$, (d) $R(\omega)$, (e) $\alpha(\omega)$ and (f) $L(\omega)$ of TaP for [100] and [001] electric field polarizations.

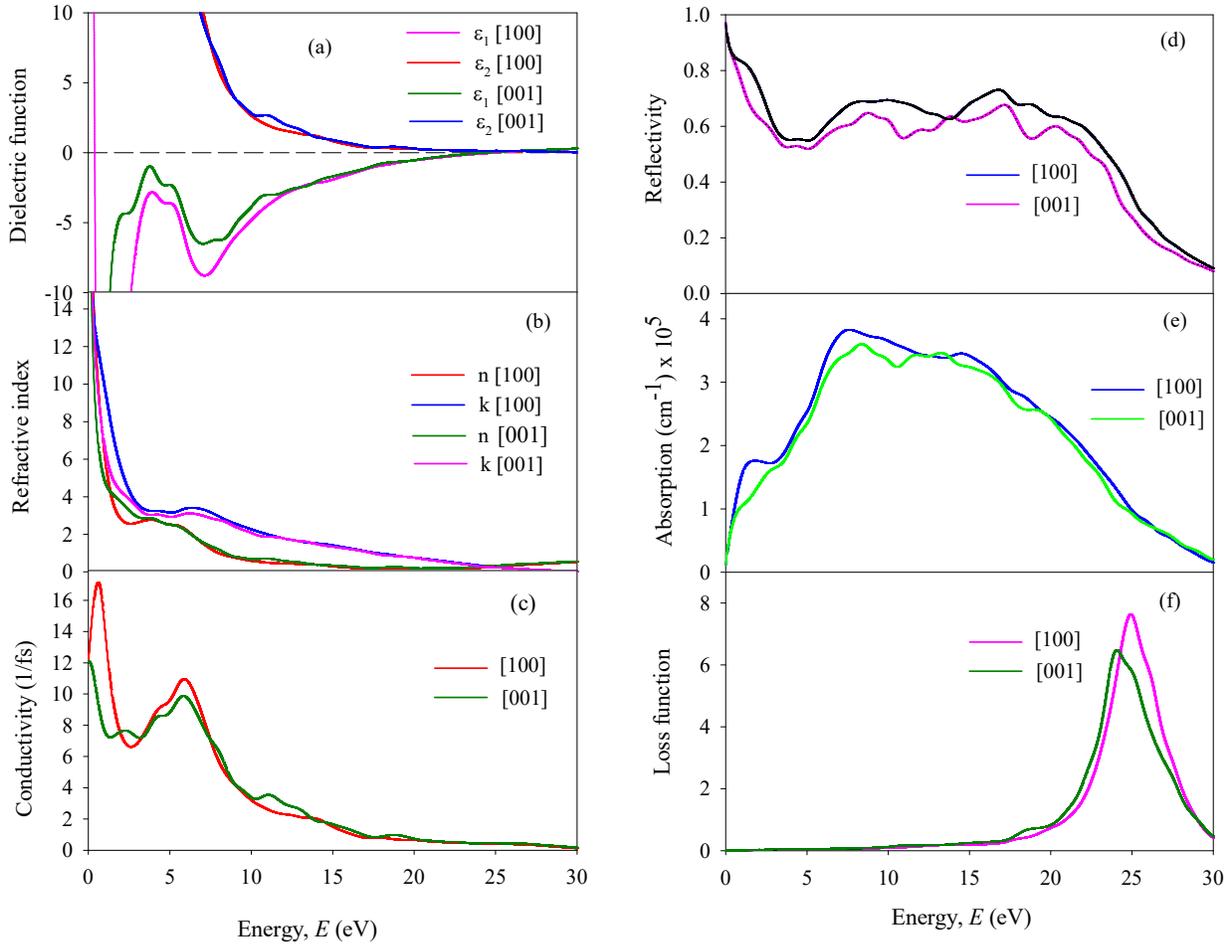

**Figure 6:** (a) $\varepsilon_1(\omega)$ and $\varepsilon_2(\omega)$, (b) $n(\omega)$ and $k(\omega)$, (c) $\sigma(\omega)$, (d) $R(\omega)$, (e) $\alpha(\omega)$ and (f) $L(\omega)$ of TaAs for [100] and [001] electric field polarizations.

## 4. Discussion and conclusions

A detailed analysis of structural, elastic, electronic, acoustic, thermal, bonding, optical and anisotropic properties of TaP and TaAs topological Weyl semimetals have been performed using the first-principles calculations. The magnitudes of the various elastic constants and moduli indicate that the mechanical failure mode of TaP and TaAs will be controlled by the shear deformation. The positive Cauchy pressure implies that TaP and TaAs ductile. On the other hand, Poisson's ratios of TaP and TaAs suggest that both the compounds should be brittle in nature. The possible source of ambiguity can be traced back to the Pugh's ratio which lies very close to the boundary value differentiating between ductile and brittle behaviors. The charge density distribution and bond population analyses indicate that both covalent and ionic bondings are present in TaP and TaAs. This is further supported by the PDOS features, where significant hybridization between Ta $5d$ and P/As $3p$ electronic orbitals are found. This indicates towards a tendency to the formation of covalent bonding between these atomic species in TaP and TaAs.

Covalent bonding with directional character tends to make both the materials brittle. Mechanical strength in both TaP and TaAs is mainly dominated by bond bending contribution. Both the compounds show moderately anisotropic charge density distribution. Low values of the universal log-Euclidean index ($A^L$) imply that the bonding strengths in TaP and TaAs are fairly isotropic along different directions within the crystal and the compounds do not show layered characteristics. TaP is significantly harder than TaAs but TaAs shows better machinability compared to TaP. It is instructive to note that the machinability indices of Ta$X$ ($X$ = P, As) are comparable to widely mentioned MAX phase nanolaminates and other layered ternary and binary compounds including topological systems [5, 6, 123-128]. The high hardness of TaP is quite remarkable and exceeds that for MAX phase compounds and related systems [123-131]. The high bond hardness of TaP found from bond hardness calculations supports the findings from the analysis of elastic constants and moduli of this WSM.

The Debye temperature and melting temperature of TaP is lower than TaAs. The low estimated value of $\Theta_D$ implies that lattice thermal conductivity of TaP and TaAs are expected to be low as well. The minimum thermal conductivity of TaP is lower than TaAs. The lower values of EDOS at the Fermi level of TaP compared to TaAs also predict that TaP will exhibit relatively low overall electronic thermal conductivity at all temperatures. Remarkably high hardness and relatively low Debye temperature of TaP makes this WSM particularly interesting with respect to its thermo-physical properties. This is a useful combination which can make TaP an efficient thermal barrier coating material. Besides, both Ta$X$ ($X$ = P, As) have high melting temperature, making them suitable for high-$T$ applications.

The electronic band structure reveals clear semimetallic character with low and almost identical values of TDOS at the Fermi level for Ta$X$ ($X$ = P, As). The Fermi level lies quite close to the pseudogap minima of both the compounds. Linear dispersion features characteristics of WSMs are seen in the band structure.

The optical constants spectra of both TaP and TaAs show close agreement with the underlying electronic band structure. The absorption coefficients of TaP and TaAs are very high in the ultraviolet region compared to other topological electronic systems [5, 6]. The refractive indices of both the compounds are also very high in the infrared-visible region. The high refractive indices of TaP and TaAs in the infrared and visible region indicate that both the compounds will be a good candidate for applications in the optimization of various display systems. The compounds under study also exhibit almost non-selective and high reflectivity over a wide range of photon energy in the ultraviolet part of the electromagnetic spectrum. Both TaP and TaAs have very high reflectivity for spectral regions covering the infrared to near-ultraviolet radiations and can be used as an efficient solar beam reflector. Furthermore, both the compounds can be utilized to confine thermal energy and to minimize heat loss to the environment. All these features hold significant promise for potential applications of TaP and TaAs.

To summarize, we have studied the elastic, mechanical, bonding, acoustic, thermal, bulk electronic and optoelectronic properties of tetragonal topological Weyl semimetals TaP and TaAs in this paper. The elastic, bonding, acoustic, thermal and optoelectronic properties are studied in depth for the first time. Both the compounds possess several attractive mechanical, thermal and optoelectronic features which are suitable for engineering and device applications. We hope that the results obtained here will inspire researchers to investigate these interesting Weyl semimetals in further details in future, both theoretically and experimentally.

**Acknowledgements**

S. H. N. acknowledges the research grant (1151/5/52/RU/Science-07/19-20) from the Faculty of Science, University of Rajshahi, Bangladesh, which partly supported this work.

**Data availability**

The data sets generated and/or analyzed in this study are available from the corresponding author on reasonable request.

# References

1. H. M. Weng, X. Dai, Z. Fang, MRS Bull. 39 (2014)849.
2. M. Wada, S. Murakami, F. Freimuth, G. Bihlmayer, Phys. Rev. B 83 (2011) 121310(R).
3. Z. Wang, Y. Sun, X. Q. Chen, C. Franchini, G. Xu, H. M.Weng, X. Dai, Z. Fang, Phys. Rev. B85 (2012) 195320.
4. R. Yu, H. M. Weng, Z. Fang, X. Dai, X. Hu, Phys. Rev. Lett. 115 (2015) 036807.
5. M. I. Naher, S. H. Naqib, J. Alloys. Compd. 829 (2020) 154509.
6. B. R. Rano, I. M. Syed, S. H. Naqib, J. Alloys. Compd. 829(2020) 154522.
7. Z. Wang, D.Gresch, A. A. Soluyanov, W.Xie, S. Kushwaha, X. Dai, M. Troyer, R. J. Cava, B. A.Bernevig, Phys. Rev. Lett. 117 (2016) 056805.
8. S.W. Boettcher, J. M. Spurgeon *et al.*, Science 347 (2015)294.
9. L. Fu, C. L. Kane, Phys. Rev. Lett. 100 (2008) 096407.
10. L. Fu, E. Berg, Phys. Rev. Lett.105 (2010) 097001.
11. P. Hosur, X. Qi, Comptes Rendus Physique 14 (2013) 857.
12. X. Wan, A.Vishwanath, S. Y. Savrasov, Phys. Rev. Lett. 108 (2012) 146601.
13. Z. Wang, S. C. Zhang, Phys. Rev. B 87 (2013) 161107.
14. H. Weyl, Z. Phys. 56 (1929) 330.
15. P. A. M. Dirac, Proc. R. Soc. London A117 (1928) 610.
16. P. B. Pal, Am. J. Phys.79 (2011) 485.
17. F. Arnold, C. Shekhar, S.Wu *et al.*, Nature Communication 7 (2016) 11615; DOI: 10.1038/ncomms11615.
18. K. Landsteiner, Phys. Rev. B 89 (2014) 075124.
19. P. Hosur, Phys. Rev. B 86 (2012) 195102.
20. A. A. Zyuzin, A. A. Burkov, Phys. Rev. B 86 (2012) 115133.

21. J. Hu, S. Y. Xu, N. Ni, Z.Mao, Annu. Rev. Mater. Res. 49 (2019) 207. https://doi.org/10.1146/annurev-matsci-070218-010023.
22. S. M. Huang, S. Y. Xu, I. Belopolski, C. C. Lee, G. Chang, B. Wang, N. Alidoust, G. Bian, M. Neupane, C. Zhang, S. Jia, A. Bansil, H. Lin, M. Z. Hasan, Nat. Commun. 6 (2015) 7373.
23. H. M. Weng, C. Fang, Z. Fang, B. A. Bernevig, X. Dai, Phys. Rev. X 5 (2015) 011029.
24. Y. Xu, I. Belopolski, D. S. Sanchez *et al.*, Sci. Adv.1 (2015) e1501092.
25. N. Xu, H. M. Weng, B. Q. Lv, C. E. Matt *et al.*, Nat. Commun. 7 (2016) 11006.
26. H. Inoue, A. Gyenis, Z. Wang*et al.*, Science 351 (2016) 1184.
27. B. Q. Lv, H. M. Weng, B. B. Fu, X. P. Wang, H. Miao, J. Ma, P. Richard, X. C. Huang, L. X. Zhao, G. F. Chen, Z. Fang, X. Dai, T. Qian, H. Ding, Phys. Rev. X 5 (2015) 031013.
28. B. Q. Lv, N. Xu, H. M. Weng *et al.*, Nat. Phys.11 (2015) 724.
29. Y. Li, Y. Zhou, Z. Guo, X. Chen, P. Lu, X. Wang, C. An, Y. Zhou, J.Xing, G. Du, X. Zhu, H. Yang, J. Sun, Z. Yang, Y. Zhang, H. H.Wen, npj Quantum Materials 2 (2017) 66.
30. Y. Zhou, P. Lu, Y. Du, X. Zhu, G. Zhang, R. Zhang, D. Shao, X. Chen,X. Wang, M. Tian, J. Sun, X. Wan, Z. Yang, W. Yang, Y. Zhang, D. Xing, Phys. Rev. Lett. 117 (2016) 146402.
31. Z. P. Guo, P. C. Lu, T. Chen, J. F. Wu, J. Sun, D. Y. Xing, Sci. China-Phys. Mech. Astron. 61 (2018) 038211.
32. H. Lu, S. Jia, Front. Phys. 12 (2017) 12721.
33. D. Chang, Y. Liu, F. Rao, F. Wang, Q. Sun,Y. Jia, Phys. Chem. Chem. Phys. 18 (2016) 14503.
34. J. Buckeridge, D. Jevdokimovs, C. R. A. Catlow, A. A. Sokol, Phys. Rev. B 93 (2016) 125205.
35. L. Liu, Z. Q. Wang, C. E. Hu, Y. Cheng,G. F. Ji, Solid State Communications 263 (2017) 10.
36. Y. Sun, S. C. Wu, B. Yan, Phys. Rev. B 92 (2015) 115428.
37. M. Dadsetani, A. Ebrahimian, Journal of Electronic Materials 45 (2016) 5867.
38. Y. Sun,Y. Zhang,C.Felser,B. Yan, Phys. Rev. Lett. 117 (2016) 146403.
39. M. Besser, R. D. Reis, F. R. Fan, M. O. Ajeesh *et al.*, Phys. Rev. Materials 3 (2019) 044201.
40. D. Sapkota, R. Mukherjee, D. Mandrus, Crystals 6 (2016) 160.
41. M. D. Segall, P. J. D. Lindan, M. J. Probert, C. J. Pickard, P. J. Harsnip, S. J. Clark, M. C. Payne, J. Phys.: Condens. Matter 14 (2002) 2717.
42. M. C. Payne, M. P. Teter, D. C. Allan, T. A. Arias, J. D. Joannopoulos, Rev. Mod. Phys. 64 (1992) 1045.
43. Materials studio CASTEP manual © Accelrys2010. http://www.tcm.phy.cam.ac.uk/castep/documentation/WebHelp/CASTEP.html
44. J. P. Perdew, K. Burke, M. Ernzerhof, Phys. Rev. Lett. 77 (1996) 3865.
45. D. Vanderbilt, Phys. Rev. B 41 (1990) 7892.
46. T. H. Fischer, J. Almlof, J. Phys. Chem. 96 (1992) 9768.
47. H. J. Monkhorst, J. D. Pack, Phys. Rev. B 13 (1976)5188.
48. G. P. Francis, M. C. Payne, J. Phys.: Condens. Matter 2 (1990) 4395.
49. O. H. Nielsen, R. M. Martin, Phys. Rev. Lett. 50 (1983) 697.
50. J. P. Watt, J. Appl. Phys. 50 (1979) 6290.
51. J. P. Watt, L. Peselnick, J. Appl. Phys. 51 (1980) 1525.
52. S. Saha, T. P. Sinha, A. Mookerjee, Phys. Rev. B 62(2000) 8828.

53. R. S. Mulliken, J. Chem. Phys. 23 (1955) 1833.
54. D. Sanchez-Portal, E. Artacho, J. M. Soler, Solid State Commun. 95 (1995) 685.
55. M. D. Segall, R. Shah, C. J. Pickard, M. C. Payne, Phys. Rev. B 54 (1996) 16317.
56. J. O. Willerstrom, J. Less-Common Met.99 (1984) 273.
57. C. C. Lee, S. Y. Xu, S. M. Huang, D. S. Sanchez, I. Belopolski, G. Q. Chang, G. Bian, N. Alidoust, H. Zheng, M. Neupane, B. K. Wang, A. Bansil, M. Z. Hasan, H. Lin, Phys. Rev. B92 (2015) 235104.
58. D. C. Wallace, *Thermodynamics of Crystal*, Wiley, New York, 1972.
59. L. Kleinman, Phys. Rev. 128 (1962) 2614.
60. W. A. Harrison, *Electronic Structure and the Properties of Solids* (San Francisco: Freeman), 1980
61. M. Jamala, S. J. Asadabadi, I. Ahmad, H. A. R. Aliabad, Computational Materials Science 95 (2014) 592.
62. A. Gueddouh, B. Bentria, I. Lefkaier, J. Magn. Magn. Mater. 406 (2016) 192.
63. M. Mattesini, R. Ahuja, B. Johansson, Phys. Rev. B 68 (2003) 184108.
64. P. Ravindran, L. Fast, P. Korzhavyi, B. Johansson, J. App. Phys. 84 (1998) 4891.
65. C. Kittel, *Introduction to Solid State Physics*, Eighth Ed., John Wiley & Sons Inc., 2005.
66. A. Yildirim, H. Koc, E. Deligoz, Chin. Phys. B 21 (2012) 037101.
67. M. Rajagopalan, S.P. Kumar, R. Anuthama, Phys. B 405 (2010) 1817.
68. W. Kim, J. Mater. Chem. C 3 (2015) 10336.
69. S. F. Pugh, Philos. Mag. 45 (1954) 43.
70. V. V. Bannikov, I.R. Shein, A.L. Ivanovskii, Physica B 405 (2010) 4615.
71. Z. Yang, D. Shi, B. Wen, R. Melnik, S. Yao, T. Li, Journal of Solid State Chemistry 183 (2010) 136.
72. G. N. Greaves, A. L. Greer, R. S. Lakes, T. Rouxel, Nature Materials 10 (2011) 823.
73. P. H. Mott, J. R. Dorgan, C. M. Roland, J. Sound Vib. 312 (2008) 572.
74. H. Fu, D. Li, F. Peng, T. Gao, X. Cheng, Comput. Mater. Sci. 44 (2008) 774.
75. A. Simunek, Phys. Rev. B 75 (2007) 172108.
76. W. Feng, S. Cui, Can. J. Phys. 92 (2014) 1652.
77. D. G. Pettifor, Mater. Sci. Technol. 8 (1992) 345.
78. M. E. Eberhart, T. E. Jones, Phys. Rev. B 86 (2012) 134106.
79. Z. Sun, D. Music, R. Ahuja, J.M. Schneider, Phys. Rev. B 71 (2005) 193402.
80. L. Vitos, P.A. Korzhavyi, B. Johansson, Nature Mater. 2 (2003) 25.
81. R.C. Lincoln, K.M. Koliwad, P.B. Ghate, phys. Rev. 157 (1967) 463.
82. K. J. Puttlitz, K.A. Stalter, Handbook of Lead-Free Solder Technology for Microelectronic Assemblies (New York: Springer) (2005) p. 98.
83. M. J. Phasha, P. E. Ngoepe, H. R. Chauke, D. G. Pettifor, D. Nguyen-Mann, Intermetallics 18 (2010) 2083.
84. X. P. Gao, Y. H. Jiang, R. Zhou, J. Feng, J. Alloys Compd. 587 (2014) 819.
85. C. M. Kube, AIP Adv. 6 (2016) 095209.

86. V.Arsigny, P. Fillard, X. Pennec and N. Ayache, Fast and simple calculus on tensors in the logeuclidean framework, in Image Computing and Computer-Assisted Intervention, Lecture Notes in Computer Science, Eds. J. Duncan and G. Gerig (Springer-Verlag, 2005), Vol. 3749, pp. 115–122, ISBN 9783540293262.
87. C. M. Kube, M. D. Jong, J. Appl. Phys. 120 (2016)165105.
88. S.Ranganathan, M. Ostoja-Starzewski, Phys. Rev. Lett. 101 (2008) 055504.
89. D. H. Chung, W.R. Buessem, in *Anisotropy in Single Crystal Refractory Compound*, Edited by F. W. Vahldiek and S. A. Mersol, Vol. 2 (Plenum press, New York, 1968).
90. D. H. Chung, W. R. Buessem, J. Appl. Phys. 38 (1967) 2010.
91. V. Milman, M. C. Warren, J. Phys.: Condens. Matter 13 (2001) 5585.
92. P. Ravindran, L. Fast, P. A. Korzhavyi, B. Johansson, J. Appl. Phys. 84 (1998) 4891.
93. E. Schreiber, O. L. Anderson, N. Soga, *Elastic Constants and Their Measurements* (McGrawHill, New York, 1973).
94. M. F. Ashby, P. J. Ferreira, and D. L. Schodek (2009). Material Classes, Structure, and Properties. Nanomaterials, Nanotechnologies and Design, p-143. doi:10.1016/b978-0-7506-8149-0.00006-4.
95. N. H. Fletcher and T. D. Rossing, The physics of musical instruments, Springer-Verlag, 1991.
96. T. Besara, D. A. Rhodes, K. W. Chen, S. Das *et al.*, Phys. Rev. B 93 (2016) 245152.
97. J. Yang, M. Shahid, C. L. Wan, F. Jing, W. Pan, J. Eur. Ceram. Soc.37 (2017) 689.
98. W. L. McMillan, Phys. Rev. 167 (1968) 331.
99. M. E. Fine, L. D. Brown, H. L. Marcus, Scr. Metall. 18 (1984) 951.
100. D. R. Clarke, Surf. Coat. Technol. 163 (2003) 67.
101. D. G. Cahill, S. K. Watson, R. O. Pohl, Phys. Rev. B 46 (1992) 6131.
102. J. H. Xu, T. Oguchi, A.J. Freeman, Phys. Rev. B 36 (1987) 4186.
103. T. Hong, T.J. Watson-Yang, A.J. Freeman, T. Oguchi, J. H. Xu, Phys. Rev. B 41 (1990) 12462.
104. A. Pasturel, C. Colinet, P. Hicter, Physica B 132 (1985) 177.
105. I. Galanakis, P. Mavropoulous, J. Phys.: Condens. Matter 19 (2007) 315213.
106. C. D. Gelatt *et al.*, Phys. Rev. B 27(1983) 2007.
107. A. Pasturel, C. Colinet, P. Hicter, Physica B 132 (1985) 177.
108. W. Lin, J. H. Xu, A. J. Freeman, Phys. Rev. B 45 (1992) 10863.
109. V. L. Moruzzi, P. Oelhafen, A. R. Williams, Phys. Rev. B 27 (1983) 2049.
110. K. H. Bennemann, J. W. Garland, in Superconductivity in d- and f- Band Metals, edited by D. H. Douglas, AIP Conf. Proc. No. 4, Edited by D. H. Douglass (AIP, New York, 1972), p.103.
111. N. E. Christensen, D. L. Novikov, Phys. Rev. B 73 (2006) 224508.
112. R. S. Mulliken, J. Chem. Phys. 23 (1955) 1833.
113. R. C. D. Richardson, Wear 10 (1967) 291.
114. V. V. Brazhkin, A. G. Lyapin, R. J. Hemley, Philos. Mag. A 82(2002) 231.
115. F. Birch, J. Geophys. Res. 83 (1978) 1257.

116. F. M. Gao, Phys. Rev. B 73 (2006) 132104.
117. F.M. Gao, J. L. He, E. D. Wu, S. M. Liu, D. L. Yu, D. C. Li, S. Y. Zhang, Y. J. Tian, Phys. Rev. Lett. 91 (2003) 015502.
118. R. D. Harcourt, J. Phys. B 7 (1974) L41.
119. A. H. Reshak, V. V. Atuchin, S. Auluck, I. V. Kityk, J. Phys. Condens. Matter 20 (2008) 325234.
120. S. Li, R. Ahuja, M. W. Barsoum, P. Jena, B. Johansson, Appl. Phys. Lett. 92 (2008) 221907.
121. H. Wang, Y. Chen, Y. Kaneta, S. Iwata, J. Alloys Compd. 491 (2010) 550.
122. S. Shaha, T. P. Sinha, A. Mookarjee, Phys. Rev. B 62 (2000) 8828.
123. F. Parvin, S. H. Naqib, J. Alloys Compd. 780 (2019) 452–460. https://doi.org/10.1016/j.jallcom.2018.12.021.
124. M. M. Hossain and S. H. Naqib, Molecular Physics (2019). https://doi.org/10.1080/00268976.2019.1609706.
125. Md. Maruf Mridha, S. H. Naqib, (2020) arXiv:2003.14146.
126. M. A. Hadi, N. Kelaidis, S. H. Naqib, A. Chroneos, A. K. M. A. Islam, J. Phys. Chem. Solids 129 (2019) 162.
127. F. Sultana, M. M. Uddin, M. A. Ali, M. M. Hossain, S. H. Naqib, A. K. M.A. Islam, Results in Physics 11 (2018) 869.
128. M. A. Hadi, U. Monira, A. Chroneos, S. H. Naqib, A. K. M. A. Islam, N. Kelaidis, R. V. Vovk, J. Phys. Chem. Solids 132 (2019) 38.
129. F. Parvin and S. H. Naqib, Chin. Phys. B 26 (2017) 106201.
130. M. A. Ali, M. A. Hadi, M. M. Hossain, S. H. Naqib, and A. K. M. A. Islam, Physica Status Solidi: B (2017). https://doi.org/10.1002/pssb.201700010.
131. M. I. Naher, F. Parvin, A. K. M. A. Islam, S. H. Naqib, Eur. Phys. J. B (2018). https://doi.org/10.1140/epjb/e2018-90388-9.

## Author Contributions

M. I. N. performed the theoretical calculations, contributed in the analysis and wrote the draft manuscript. S. H. N. designed and supervised the project, analyzed the results and finalized the manuscript. Both the authors reviewed the manuscript.

## Additional Information

## Competing Interests

The authors declare no competing interests.